\documentclass[twoside,twocolumn,9pt]{article}
\usepackage{extsizes}
\usepackage[super,sort&compress,comma]{natbib} 
\usepackage[version=3]{mhchem}
\usepackage[left=1.5cm, right=1.5cm, top=1.785cm, bottom=2.0cm]{geometry}
\usepackage{balance}
\usepackage{times,mathptmx}
\usepackage{sectsty}
\usepackage{graphicx} 
\usepackage{lastpage}
\usepackage[format=plain,justification=justified,singlelinecheck=false,font={stretch=1.125,small,sf},labelfont=bf,labelsep=space]{caption}
\usepackage{float}
\usepackage{fancyhdr}
\usepackage{fnpos}
\usepackage[english]{babel}
\addto{\captionsenglish}{%
  
}
\usepackage{array}
\usepackage{droidsans}
\usepackage{charter}
\usepackage[T1]{fontenc}
\usepackage[usenames,dvipsnames]{xcolor}
\usepackage{setspace}
\usepackage[compact]{titlesec}
\usepackage{hyperref}

\usepackage{epstopdf}

\definecolor{cream}{RGB}{222,217,201}

\begin{document}

\pagestyle{fancy}
\thispagestyle{plain}
\fancypagestyle{plain}{

\fancyhead[C]{\includegraphics[width=18.5cm]{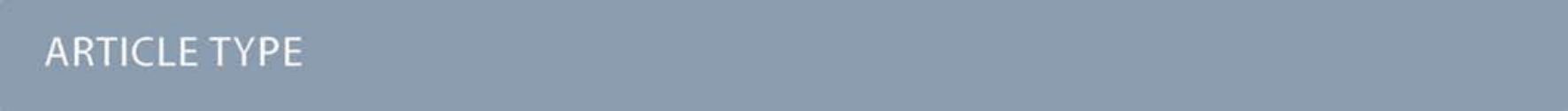}}
\fancyhead[L]{\hspace{0cm}\vspace{1.5cm}\includegraphics[height=30pt]{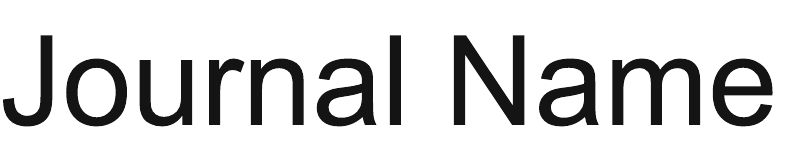}}
\fancyhead[R]{\hspace{0cm}\vspace{1.7cm}\includegraphics[height=55pt]{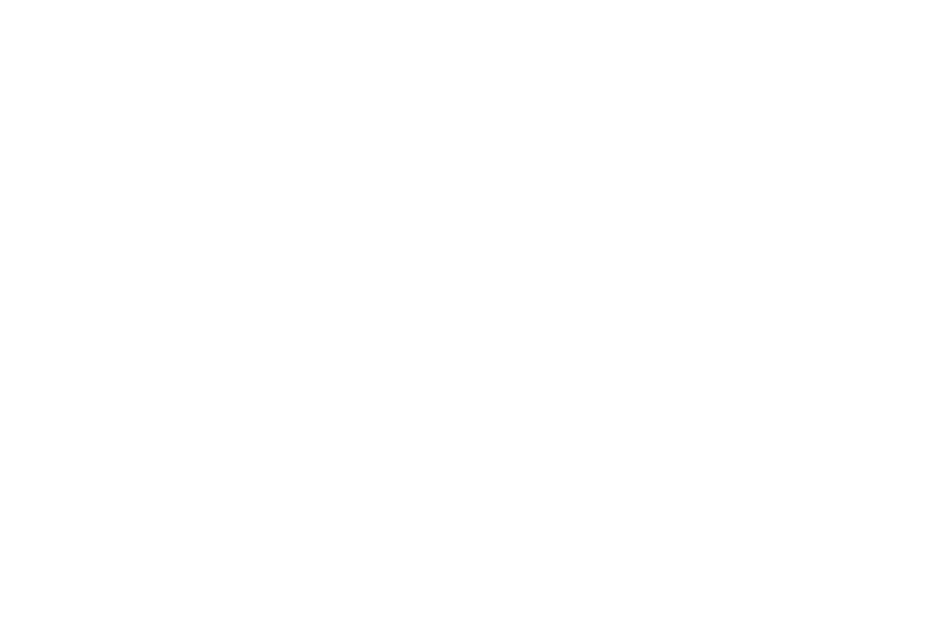}}
\renewcommand{\headrulewidth}{0pt}
}

\makeFNbottom
\makeatletter
\renewcommand\LARGE{\@setfontsize\LARGE{15pt}{17}}
\renewcommand\Large{\@setfontsize\Large{12pt}{14}}
\renewcommand\large{\@setfontsize\large{10pt}{12}}
\renewcommand\footnotesize{\@setfontsize\footnotesize{7pt}{10}}
\makeatother

\renewcommand{\thefootnote}{\fnsymbol{footnote}}
\renewcommand\footnoterule{\vspace*{1pt}%
\color{cream}\hrule width 3.5in height 0.4pt \color{black}\vspace*{5pt}} 
\setcounter{secnumdepth}{5}

\makeatletter 
\renewcommand\@biblabel[1]{#1}            
\renewcommand\@makefntext[1]%
{\noindent\makebox[0pt][r]{\@thefnmark\,}#1}
\makeatother 
\renewcommand{\figurename}{\small{Fig.}~}
\sectionfont{\sffamily\Large}
\subsectionfont{\normalsize}
\subsubsectionfont{\bf}
\setstretch{1.125} 
\setlength{\skip\footins}{0.8cm}
\setlength{\footnotesep}{0.25cm}
\setlength{\jot}{10pt}
\titlespacing*{\section}{0pt}{4pt}{4pt}
\titlespacing*{\subsection}{0pt}{15pt}{1pt}

\newcommand{\ben}{\begin{equation}}
\newcommand{\een}{\end{equation}}
\newcommand{\bea}{\begin{eqnarray}}
\newcommand{\eea}{\end{eqnarray}}


\fancyfoot{}
\fancyfoot[LO,RE]{\vspace{-7.1pt}\includegraphics[height=9pt]{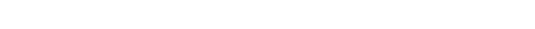}}
\fancyfoot[CO]{\vspace{-7.1pt}\hspace{13.2cm}\includegraphics{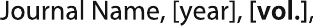}}
\fancyfoot[CE]{\vspace{-7.2pt}\hspace{-14.2cm}\includegraphics{RF}}
\fancyfoot[RO]{\footnotesize{\sffamily{1--\pageref{LastPage} ~\textbar  \hspace{2pt}\thepage}}}
\fancyfoot[LE]{\footnotesize{\sffamily{\thepage~\textbar\hspace{3.45cm} 1--\pageref{LastPage}}}}
\fancyhead{}
\renewcommand{\headrulewidth}{0pt} 
\renewcommand{\footrulewidth}{0pt}
\setlength{\arrayrulewidth}{1pt}
\setlength{\columnsep}{6.5mm}
\setlength\bibsep{1pt}

\makeatletter 
\newlength{\figrulesep} 
\setlength{\figrulesep}{0.5\textfloatsep} 

\newcommand{\topfigrule}{\vspace*{-1pt}%
\noindent{\color{cream}\rule[-\figrulesep]{\columnwidth}{1.5pt}} }

\newcommand{\botfigrule}{\vspace*{-2pt}%
\noindent{\color{cream}\rule[\figrulesep]{\columnwidth}{1.5pt}} }

\newcommand{\dblfigrule}{\vspace*{-1pt}%
\noindent{\color{cream}\rule[-\figrulesep]{\textwidth}{1.5pt}} }

\makeatother

\twocolumn[
  \begin{@twocolumnfalse}
\vspace{3cm}
\sffamily
\begin{tabular}{m{4.5cm} p{13.5cm} }

\includegraphics{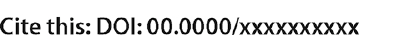} & \noindent\LARGE{\textbf{Excitons in two-dimensional atomic layer materials from time-dependent density functional theory: mono-layer and bi-layer hexagonal boron nitride and transition-metal dichalcogenides
}} \\
\vspace{0.3cm} & \vspace{0.3cm} \\

 & \noindent\large{Yasumitsu Suzuki$^{\ast}$\textit{$^{a}$} and Kazuyuki Watanabe\textit{$^{a}$}} \\

\includegraphics{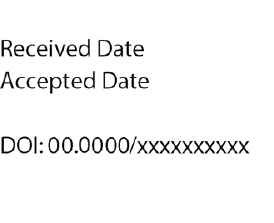} & \noindent\normalsize{Time-dependent density functional theory (TDDFT) has been applied 
to the calculation of absorption spectra for two-dimensional atomic layer materials;
mono-layer and bi-layer hexagonal boron nitride (h-BN) and mono-layer transition metal dichalcogenides, MoS$_2$ and MoSe$_2$.
We reveal that the character of the first bright exciton state of bi-layer h-BN 
is dependent on the layer stacking type through the use of many-body perturbation theory (MBPT) calculations,
i.e., the electron and hole in the AA$'$ stacking are present in the same layer (intralayer exciton) while the A$'$B stacking exhibits an interlayer exciton.
We demonstrate that the TDDFT approach with the meta-generalized gradient approximation to the exchange-correlation (XC) potential and 
the Bootstrap XC kernel can capture the characteristics of the absorption peaks that correspond to these excitons  
without computationally heavy GW and Bethe-Salpeter equation calculations. 
We also show that the TDDFT method 
can capture the qualitative features of
the absorption spectra for 
mono-layer transition metal dichalcogenides, MoS$_2$ and MoSe$_2$,
although the exciton binding energies are underestimated.
This study elucidates the usefulness of the TDDFT approach
for the qualitative investigation of the optical properties of  
two-dimensional atomic layer materials.} \\

\end{tabular}

 \end{@twocolumnfalse} \vspace{0.6cm}

  ]

\renewcommand*\rmdefault{bch}\normalfont\upshape
\rmfamily
\section*{}
\vspace{-1cm}


\footnotetext{\textit{$^{a}$~Department of Physics, Tokyo University of Science, 1-3 Kagurazaka, Shinjuku-ku, Tokyo 162-8601, Japan.}}



\section{Introduction}\label{s:intro}

Two-dimensional (2D) atomic layer materials have been gathering more and more attention since an efficient production method for the representative 2D material, graphene,
 was discovered in 2004~\cite{graphene}.
Many novel properties and physics that appear in these 2D materials due to their distinct electronic structures and the quantum confinement effect have been experimentally and theoretically discovered, such as
the quantum Hall effect~\cite{hall1,hall2}, tightly bound trions~\cite{trion1,trion2,trion3}, and the valley degrees of freedom~\cite{valley1,valley2,valley3}, just to name a few. 
Various 2D materials have been synthesized, including metals (e.g., graphene~\cite{graphene}), semiconductors (e.g., mono- and few-layer transition metal dichalcogenides (TMDCs)~\cite{tmdc1,tmdc2,tmdc3}), and insulators (e.g., mono- and few-layer hexagonal boron nitride (h-BN)~\cite{hBN1}). These materials have been extensively studied towards their application to nanoscale-thermoelectronic ~\cite{thermo0,thermo1,thermo2,thermo3} and optoelectronic devices~\cite{optel1,optel2,optel3}, such as light-harvesting devices~\cite{harvest1,harvest2,harvest3,harvest4}, photo-detectors~\cite{detect1,detect2}, and light-emitting devices~\cite{emit1,emit2,emit3,emit4}.
These 2D materials can be combined by van der Waals (vdW) forces as building blocks to produce the vdW 
heterostructures~\cite{vdW1,vdW2,vdW3,vdW4,vdW5}, which can be high-performance optoelectronic devices~\cite{opt1,opt2,opt3,opt4,opt5,opt6,opt7,opt8,opt9,opt10}.
Atoms or molecules can also be intercalated or adsorbed into these 2D materials as chemical dopants to modify the number of carriers, and to control the absorbance and photoluminescence properties~\cite{dope1, dope2,dope3}.
   
When considering the application of these 2D materials to nano-optoelectronic devices,
a first-principles analysis of their optical properties plays a crucial role.
If the absorption spectrum (i.e., the imaginary part of macroscopic dielectric function) of the materials could be simulated, then it would be possible 
to design the best combination of materials that most efficiently absorb a specific wavelength of light, for example.
However, the dielectric screening of the Coulomb interaction in many 2D materials is significantly reduced due to their atomically thin structures when compared to the bulk, and the electron and hole are strongly bound after the absorption of light, i.e., the excitonic effect is enhanced~\cite{sim2d1}. Consequently, the independent-particle transition picture is considered to break down, and calculation based on Fermi's golden rule, or the random phase approximation (RPA)~\cite{rpa1,rpa2,mbpt1} is not adequate to calculate the absorption spectrum of these 2D materials accurately~\cite{sim2d1}. The method based on the many-body perturbation theory (MBPT)~\cite{mbpt1,mbpt2}, i.e., solving the Bethe-Salpeter equation (BSE) with the GW approximation has generally been used to 
calculate the electron self-energy (quasiparticle effect) and 
the excitonic states and the macroscopic dielectric function of 
these materials~\cite{emit4,opt5,sim2d1,bse-1,bse0,bse1,bse2,bse3,bse4,bse-m-hBN0,bse-m-hBN1,bse-m-hBN2,bse-bi-hBN1,bse-bi-hBN2}.
However, it is known that both the GW calculation and the BSE calculation require huge computational costs, and thus the application of the MBPT approach to the complex 2D systems, such as heterostructures and molecule-adsorbed systems, remains a challenge, even with the current supercomputers~\cite{opt8,opt9,bsehetero1,bsehetero2,bsehetero3}.
 
In this study, we investigate the possibility of calculating the absorption spectra of 2D materials using the method based on {\it time-dependent density functional theory (TDDFT)}~\cite{tddft1,tddft2,tddft3}.
TDDFT is a formally exact approach to the time-dependent many-body problem, and in principle it can capture the quasiparticle and excitonic effects, and give the correct spectra~\cite{tddft2}, as we briefly review these in the next section.
The TDDFT approach in the linear response regime can significantly reduce the computational cost compared to the MBPT approach. 
This is because TDDFT relies on the electron density instead of the Green's function, and only two-point response functions
are involved in the formalism, while the MBPT approach requires the four-point response functions~\cite{tddft2,fxc1,fxc2,fxc3}.  
In practice, the exchange-correlation (XC) term that incorporates all many-body effects in the theory must be approximated, and the accuracy of the calculation depends on the quality of the approximation. 
There has been a long time effort to improve the XC term to capture the excitonic effect in a
material~\cite{tddft2,fxc1,fxc2,fxc3,fxc4,fxc5,boot1,fxc6,fxc7,fxc8,fxc9,fxc92,fxc10,fxc11,fxc12,fxc13}.
Among these efforts, the Bootstrap XC kernel developed by Sharma {\it et al}.~\cite{boot1,boot2} has been reported to be effective to provide a description of the excitonic peaks in the absorption spectra
without a requirement for material-dependent parameters.
 On the other hand, the meta-generalized gradient approximation to the XC potential 
developed by Tran and Blaha (TB-mBJ potential)~\cite{tb1,tb2,tb3}
is known to give correct band gaps of materials, i.e., effectively capture the quasiparticle effect.
There have been many successful applications of the TDDFT approach to the isolated systems and the
three dimensional bulk systems~\cite{tddft2,fxc1,fxc2,fxc3,fxc4,fxc5,boot1,fxc6,fxc7,fxc8,fxc9,fxc92,fxc10,fxc11,fxc12,fxc13}.
Nevertheless, the application of the TDDFT approach to 2D materials
has been limited to date~\cite{tddft2D1,tddft2D2,tddft2D3,tddft2D4,tddft2D5,tddft2D6,tddft2D7,tddft2D8}, and there have been no studies that applied TDDFT to calculation of the excitonic peaks in the absorption spectra of 2D materials, to the best of our knowledge.
Therefore, it is unclear whether the TDDFT approach can describe the absorption peaks that correspond to the interlayer excitons where the electron and hole sit on different layers, which can emerge in multi-layer 2D systems and is important for application 
in nano-optoelectronics~\cite{opt2,opt5,opt8,opt9,opt10,bse-bi-hBN1,bse-bi-hBN2,bsehetero3}.

Here we calculate the absorption spectra of 2D materials, mono-layer and bi-layer hexagonal boron nitride (h-BN) and mono-layer TMDCs (MoS$_2$ and MoSe$_2$), using the TDDFT approach with the TB-mBJ potential~\cite{tb1,tb2,tb3} and the Bootstrap kernel~\cite{boot1,boot2}.
This is the first application of the Bootstrap kernel and TB-mBJ potential to calculation of the absorption spectra for 2D materials.
A recent study~\cite{bse-bulk-hBN} reported that the character of excitons in bulk h-BN is governed by the layer stacking arrangement.   
Here we show that the excitons in bi-layer h-BN also have a dependence on the layer stacking by the MBPT calculation, and then demonstrate that the TDDFT approach qualitatively describes the absorption peaks that correspond to these excitons.
We also show that the TDDFT approach qualitatively captures the difference in the absorption spectra between mono-layer MoS$_2$ and MoSe$_2$
reported from experimental measurements~\cite{tmdcexp}.

\section{Methods}\label{s:method}

\subsection{Formalism} \label{ss:formalism}
In this section we describe the TDDFT approach used in this study.
We first review how TDDFT is connected with MBPT and the rigorous form of the XC kernel~\cite{tddft2} to remind the reader that TDDFT in principle can describe the excitonic peak energies.
We then show 
that the Bootstrap kernel~\cite{boot1,boot2} can be derived from the exact XC kernel,
presenting the approximations included in the derivation. 
That is, we demonstrate the features of the exact XC kernel that are captured by the Bootstrap kernel
with the aim to represent the question whether or not these features are effective to describe the excitonic peaks of 2D materials.
 
In the linear response TDDFT regime, the response function $\chi$ is given by the following
Dyson-like equation~\cite{tddft2} (atomic units are used unless stated otherwise):
\ben
\begin{split}
&\chi(1,2)=\chi_s(1,2)\\
&+\int d3d4 \chi_s(1,3)[v(3,4)+f_{\rm xc}(3,4)]\chi(4,2),
\end{split}
\label{eqn:dyson}
\een
where $\chi_s$ is the response function of the non-interacting Kohn-Sham system,
$v$ is the bare Coulomb potential,
and $f_{\rm xc}$ is the XC kernel
(the numbers represent space-time arguments, e.g., $1=({\bf r}_1,t_1)$).
The dielectric function is obtained from $\chi$ 
through $\varepsilon^{-1}({\bf q},\omega)=1+v({\bf q})\chi({\bf q},\omega)$.

\begin{figure}[h]
 \centering
 \includegraphics*[width=1.0\columnwidth]{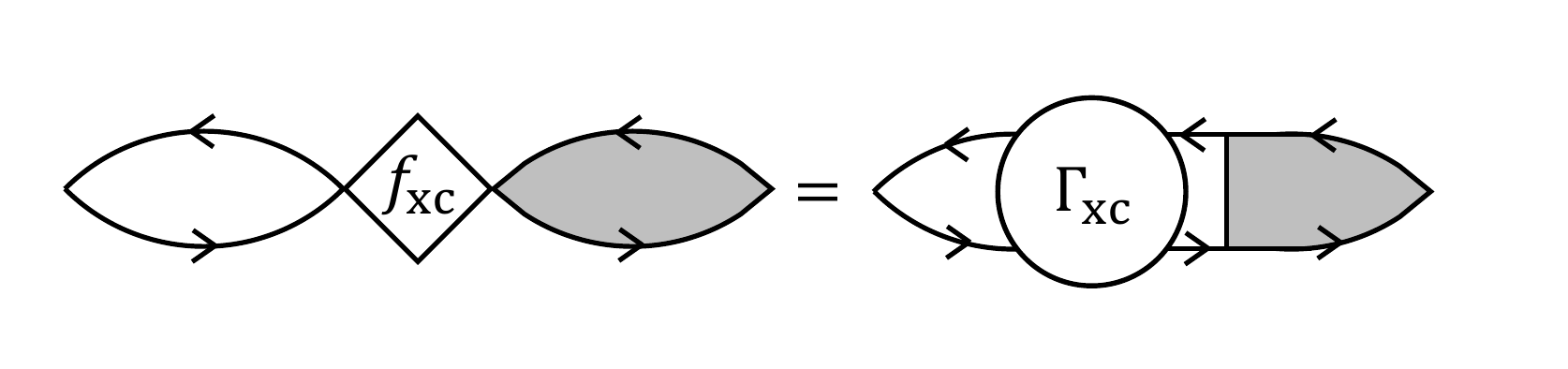}
 \caption{Diagrammatic representation of Eq.~(\ref{eqn:fxcbse}) that connects the XC kernel in TDDFT with the four-point function in MBPT.
}
 \label{Fig1}
\end{figure}

Next we present the connection between TDDFT and MBPT.
Defining the proper response function $\tilde{\chi}$ as
$\chi=\tilde{\chi}+\tilde{\chi}v\chi$ (integration is implied in the appropriate places)~\cite{tddft2}, the following expression for $f_{xc}$ is derived:
\ben
\tilde{\chi}(1,2)=\chi_s(1,2)+\int d3d4 \chi_s(1,3)f_{\rm xc}(3,4)\tilde{\chi}(4,2).
\label{eqn:properchi}
\een
In MBPT, the propagation of an electron-hole pair is described by the two-particle Green's function, or the four-point 
polarization function $L(1,2,3,4)$, which satisfies the BSE~\cite{mbpt1,mbpt2}:
\ben
\begin{split}
&L(1,2,3,4)=L_0(1,2,3,4)\\
&+\int d5d6d7d8 L_0(1,7,3,5)\Gamma(5,6,7,8)L(8,2,6,4).
\end{split}
\label{eqn:bse}
\een
$L_0$ is the non-interacting four-point polarization and 
$\Gamma$ is the two-particle scattering amplitude.
In the equal time limit, $L(1,2,1,2)$ reduces to $\chi(1,2)$, and thus the dielectric function that 
takes account of excitons is obtained once the BSE is solved.
Now with the definition of the proper four-point polarization $\tilde{L}$ as  
$L=\tilde{L}+\tilde{L}wL$ and $\tilde{L}=L_s+L_s\Gamma_{\rm xc}\tilde{L}$
(where $w(1,2,3,4)=\delta(1,3)\delta(2,4)v(1,2)$, $L_s$ is the four-point polarization of the Kohn-Sham system, and 
$\Gamma_{\rm xc}$ is the XC scattering amplitude), the XC kernel that gives
the same response function as that from the BSE is given by the following equation~\cite{tddft2}:
\ben
\begin{split}
&\int d3d4 \chi_s(1,3)f_{\rm xc}(3,4)\tilde{\chi}(4,2)\\
&=\int d3d4d5d6 L_s(1,5,1,3)\Gamma_{\rm xc}(3,4,5,6)\tilde{L}(6,2,4,2),
\end{split}
\label{eqn:fxcbse}
\een
which is represented diagrammatically in Fig.~\ref{Fig1}.

Equation~(\ref{eqn:fxcbse}) connects the XC kernel in TDDFT with the four-point function in MBPT, and this equation
should be the starting point to develop the approximation to the XC kernel.
The Bootstrap kernel~\cite{boot1,boot2} does capture certain features of the exact XC kernel~(\ref{eqn:fxcbse}).
To observe this, the partition of the XC kernel as a sum of a quasiparticle part $f_{\rm xc}^{\rm qp}$ and 
an excitonic part $f_{\rm xc}^{\rm ex}$, $f_{\rm xc}=f_{\rm xc}^{\rm qp}+f_{\rm xc}^{\rm ex}$, is first introduced with the following relations~\cite{fxcqp}:
\ben
\tilde{\chi}=\chi_{\rm qp}+\chi_{\rm qp}f_{\rm xc}^{\rm ex}\tilde{\chi},
\label{eqn:fxc_ex}
\een
\ben
\chi_{\rm qp}=\chi_s+\chi_sf_{\rm xc}^{\rm qp}\chi_{\rm qp},
\label{eqn:fxc_qp}
\een
where $\chi_{\rm qp}$ is the response function of non-interacting quasiparticles, i.e.,
$f_{\rm xc}^{\rm qp}$ and $f_{\rm xc}^{\rm ex}$ bear the quasiparticle and excitonic effect, respectively.
The equation that connects $f_{\rm xc}^{\rm ex}$ with the four-point polarization of the quasiparticle system
can now be approximated. 
Replacement of the scattering amplitude with the screened Coulomb interaction $W$~\cite{mbpt1,mbpt2}, and the replacement of $\tilde{\chi}$ and $\tilde{L}$
with the corresponding quasiparticle expressions leads to the so-called ``nanoquanta'' kernel~\cite{tddft2,fxc1,fxc2,fxc3}:
\ben
\begin{split}
&\int d3d4 \chi_{\rm qp}(1,3)f_{\rm xc}^{\rm ex}(3,4)\chi_{\rm qp}(4,2)\\
&=\int d3d4 G_{\rm qp}(1,3)G_{\rm qp}(4,1)W(3,4)G_{\rm qp}(3,2)G_{\rm qp}(2,4),
\end{split}
\label{eqn:fxcnano}
\een
where $G_{\rm qp}$ is the one-particle Green's function.
The nanoquanta kernel captures the excitonic effect through Eq.~(\ref{eqn:fxcnano}), but is
computationally as expensive as the BSE. 
One of the important features of the nanoquanta kernel is that 
it has the form of 
$f_{\rm xc}^{\rm ex}\sim \frac{\varepsilon^{-1}(0)}{q~2}\alpha$
in the optical limit (${\bf q}\to 0$)~\cite{fxc5}.
It has been reported that this ${\bf q}^{-2}$-divergence~\cite{div1,div2} in the XC kernel is crucial to take account of the excitonic effect~\cite{tddft2,fxc4,fxc5,fxc10}.

And from this aspect, the Bootstrap kernel proposed by Sharma {\it et al.}~\cite{boot1,boot2},
\ben
f_{\rm xc}^{\rm boot}=\frac{\varepsilon^{-1}({\bf q},\omega=0)}{\chi_s({\bf q},\omega=0)},
\label{eqn:fxc_boot}
\een
does capture the feature of the nanoquanta kernel because $\chi_s({\bf q}\to 0)=x_0q^2$~\cite{fxc10,div1,div2},
i.e., the Bootstrap kernel contains the long range part of the nanoquanta kernel,
without the requirement of material-dependent parameters or heavy computational cost.
In this study we investigate whether these features of the Bootstrap kernel are effective to capture the 
excitonic effect in 2D materials,
in particular the interlayer exciton, which involves 
a charge transfer that may require a non-local exchange kernel~\cite{EXX} to be accurately described.

We also explore whether the {\it full density-functional} approach can be used to predict 
the absorption spectra of 2D materials.
For this purpose, $\chi_{\rm qp}$ must also be calculated without performing heavy GW calculations~\cite{mbpt1,mbpt2}.
Here we do not solve Eq.~(\ref{eqn:fxc_qp}) with $f_{\rm xc}^{\rm qp}$, but instead $\chi_{\rm qp}$ is approximated by the Kohn-Sham response function obtained with the TB-mBJ potential functional~\cite{tb1,tb2},
which is known to yield correct band gaps and thus is expected to take account of the quasiparticle effect.
We note that here we do not use the scissor correction~\cite{scissor} that 
shifts the conduction bands, because it first requires to determine 
the value of the scissor shift from, e.g. the experimental gap or the GW calculation,
while the purpose of the present study is 
to investigate the accuracy of the full density-functional method that does not use
the experimental data or the computationally heavy MBPT calculation.

\subsection{Computational details} \label{ss:details}

The TDDFT and MBPT calculations have been performed
using the full-potential linearized augmented plane wave codes {\tt Elk}~\cite{elk} and {\tt exciting}~\cite{exciting1,exciting2,exciting3}, respectively.
In the ground-state calculation and the TDDFT calculation, a basis-set cutoff $R_{\rm MT}G_{\rm max}=7$ is used, while in the BSE calculation,
$R_{\rm MT}G_{\rm max}=6$ is used. A muffin-tin radius $R_{MT}=1.3$ bohr is adopted for all atomic species involved in the simulation.
In the MBPT calculation, 
the ground-state electron densities are computed in the framework of density functional theory using the PBE XC potential functional~\cite{pbe}. 
Sampling of the Brillouin zone is $18\times18\times1$ {\bf k}-points for the ground-state calculation, 
$10\times10\times1$ shifted {\bf k}-points for calculation of the quasiparticle correction to the Kohn-Sham eigenvalue within the G$_0$W$_0$ approximation, and $24\times24\times1$ shifted {\bf k}-points for the BSE calculation.
60 empty states are included in all G$_0$W$_0$ and BSE calculations.
In the construction of the BSE Hamiltonian, two occupied and two unoccupied bands are considered for mono-layer h-BN, and four occupied and four unoccupied bands are considered for bi-layer h-BN.
In the TDDFT calculation, the TB-mBJ potential functional~\cite{tb1} is used to obtain the approximate quasiparticle response function $\chi_{\rm qp}$, as explained in the previous section, and the Bootstrap kernel~\cite{boot1} is used to solve the Dyson-like equation~(\ref{eqn:dyson}). 
Local-field effects (LFE)~\cite{fxc10,boot2} are taken into account
((3 $|{\bf G}|$ vectors were included to
calculate the response functions), although we confirmed that the inclusion of LFE 
does not play an important role for the discussion below.
Brillouin zone sampling is $24\times24\times1$ {\bf k}-points.
Four empty bands per atom were included in our TDDFT calculations.
We checked that our results are sufficiently converged with respect to these parameters for our discussion below.

\section{Results}\label{s:result}

\subsection{Excitons in bi-layer h-BN from the MBPT approach} \label{ss:mbpt}

We first discuss the absorption spectra of bi-layer h-BN calculated with the MBPT approach.
Both bulk and a few-layer h-BNs are insulators that exhibit large excitonic peaks in the absorption spectra~\cite{,bse-m-hBN0,bse-m-hBN1,bse-m-hBN2,bse-bi-hBN1,bse-bi-hBN2,bse-bulk-hBN,bse-bulk-hBN2,bse-bulk-hBN3}. 
A recent study~\cite{bse-bulk-hBN} has reported that the characteristics of excitons in bulk h-BN are largely influenced by the layer stacking; 
for example, the 3D exciton, where the electron distribution overlaps with the hole distribution and also extends to the adjacent layers, appears only in the layer configuration where inversion symmetry is present.
The dependence of the exciton characteristics on the layer stacking arises from the difference in the electronic structure of the bands that contribute to the excitonic states~\cite{bse-bulk-hBN}.

In this subsection, we investigate whether the dependence of excitons on the layer stacking appear also in bi-layer h-BN using the MBPT calculation,
and the characteristics of excitons that dominate the onset of absorption.
We note that it is not obvious whether the similar dependence with those in bulk h-BN appear in bi-layer h-BN
because the reduction of dimensionality changes the electronic structures.
Our purpose in this study is to confirm whether or not the TDDFT method can be used to simulate the absorption spectra of 2D materials,
 in particular whether it can describe the first absorption peaks that may correspond to interlayer or intralayer excitons. 
Therefore, we must first clarify in which layer configuration the interlayer and intralayer excitons appear as the first peak in the absorption spectra of bi-layer h-BN.

To this end, we focus on two types of layer stacking: AA' and A'B configurations.
In the AA' stacking configuration, B and N atoms of the first layer lie on top of different atoms in the second layer, while
in A'B stacking, the N atoms occupy the hollow position and the B atoms are aligned on top of each other~\cite{bse-bulk-hBN}.
These two configurations are selected because they are reported to exhibit 
totally different excitonic characters in bulk-hBN~\cite{bse-bulk-hBN},
and it is expected that the difference between these configurations will also appear in the bi-layer case.

\begin{figure}[h]
\begin{center}
   \includegraphics[width=1.0\columnwidth]{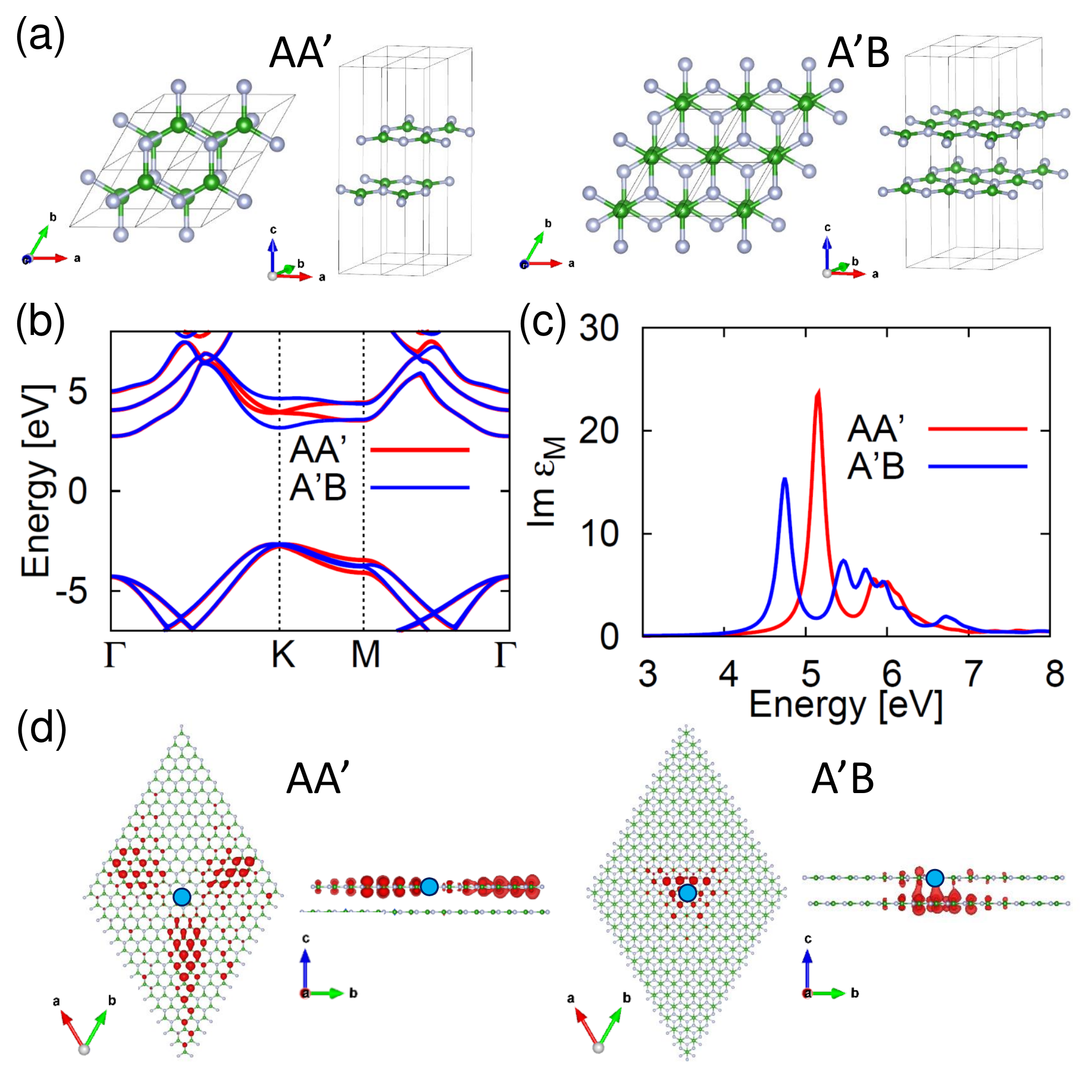}
\end{center}
\caption{\label{Fig2}
(a) Two types of layer-stacking configuration that were examined for bi-layer h-BN: AA' stacking (left) and A'B stacking (right). 
For the sake of clarity, the figures show the unit cell repeated four times.
B atoms are green and N atoms are gray.
(b) Quasiparticle band structures of two considered stacking arrangements (AA': red and A'B: blue).
(c) Optical absorption spectra for the
two considered stacking arrangements (AA': red and A'B: blue) calculated by the MBPT (G$_0$W$_0$+BSE) approach.
(d) Real-space distribution of the electron component (red isosurface) of the exciton wavefunction that gives rise to the 
first peak of the absorption spectrum (left: AA' stacking, right: A'B stacking).
Isovalues of 12\% of the maximum values are adopted.
The position of the fixed hole is indicated by a blue dot.
}
\end{figure}

Figure~\ref{Fig2}(a) shows the unit cells (repeated four times for the sake of clarity) for AA' stacking (left) and A'B stacking (right) bi-layer h-BN.
We set the in-plane lattice parameter as $2.50$ \text{\AA} for both stacking arrangements, 
while the interlayer distances are set as $3.30$ \text{\AA} for AA' stacking and as $3.24$ \text{\AA} for A'B stacking,
so that these parameters are the same as those of their bulk counterparts~\cite{bse-bulk-hBN}.
To model the 2D bi-layer system, the unit cell must have a vacuum region in the out-of-plane direction (c-direction in Fig.~\ref{Fig2}(a))
that should be sufficiently large to avoid interaction between replica images due to the periodic boundary conditions and to correctly take account of the reduction of screening. 
We set the length of the vacuum region as $18$ bohr ($\approx9.5$ \text{\AA}) for both stacking arrangements because in the previous study on mono-layer h-BN~\cite{bse-m-hBN0},
 it was demonstrated that the impact of the vacuum region length on the absorption spectrum starts to converge when the length exceeds approximately 18 bohr. Thus, we consider that 18 bohr vacuum region is sufficient for at least a primary discussion that focuses on the difference in the characteristics of excitonic peaks between different stacking systems.
 
Quasiparticle band structures calculated for these two systems by the MBPT approach (with G$_0$W$_0$ approximation) are shown in 
Fig.~\ref{Fig2}(b).
The overall band structures are similar between AA' (red) and A'B (blue) stacking. 
However, the two lowest conduction bands are degenerate at the K point in the Brillouin zone for AA' stacking, while they are energetically split for A'B stacking.
Similar degeneracy and splitting of the bands were reported for bulk h-BN case (along the 
K-H path)~\cite{bse-bulk-hBN, bse-bulk-hBN2, bse-bulk-hBN4, bse-bulk-hBN5}, 
and the mechanism should also be the same; the degeneracy of the two lowest conduction bands and the degeneracy of the two highest valence bands occur when the atoms of the same species do not lie on top of each other (such as AA' stacking), while they are split when they are on top of each other (such as A'B stacking) because of the weak electrostatic interaction between atoms of the same species.
 
The optical absorption spectra (the imaginary part of the macroscopic dielectric function) of these two systems calculated by solving the BSE are presented in Fig.~\ref{Fig2}(c) 
to show how the differences in the electronic states result in the excitonic peaks.
It is evident that there is a considerable difference between the two absorption spectra, 
i.e., the first intense excitonic peak appears at a lower energy in the spectrum for A'B stacking (blue) than that for AA' stacking (red);
 the onset of absorption is red-shifted with A'B stacking.
Similar differences in the absorption spectra between AA' and A'B stacking have been reported for bulk h-BN~\cite{bse-bulk-hBN},
and here we have shown that there is also dependence of the absorption spectra on the layer stacking
in 2D bi-layer h-BN. 

To understand the difference in the excitonic characteristics between these two different layer-stacking bi-layer h-BNs, the 
real-space distribution of the electron component of the exciton wavefunction that gives rise to the 
first peak of each absorption spectrum is plotted in Fig.~\ref{Fig2} (d) as red isosurfaces (isovalues of 12\% of the maximum values are adopted).
The position of the fixed hole is indicated by a blue dot in the figure.
For AA' stacking (left), it is clear that the electron and hole exist in the same layer, i.e., the intralayer exciton is responsible for the first absorption peak, while for the A'B exciton case (right), a major portion of the electron and hole are located on different layers, and thus it can be regarded as an interlayer exciton.
Therefore, these two different layer-stacking bi-layer h-BNs exhibit different absorption peaks that have
different excitonic characteristics. Bi-layer h-BNs thus represent ideal systems to investigate the effectiveness of the TDDFT approach to the
study of the optical properties of 2D materials, which will be clarified by
 TDDFT calculations for these two different excitonic peaks.

\subsection{TDDFT calculation of the absorption spectra of 2D materials} \label{ss:tddft}

The results of TDDFT calculations of the absorption spectra of 2D materials are presented here.
Before the results for the bi-layer h-BNs are discussed, we first present the results for mono-layer h-BN,
for which many reference data are available from previous studies~\cite{bse-m-hBN0,bse-m-hBN1,bse-m-hBN2}.
\begin{figure}[h]
\begin{center}
   \includegraphics[width=1.0\columnwidth]{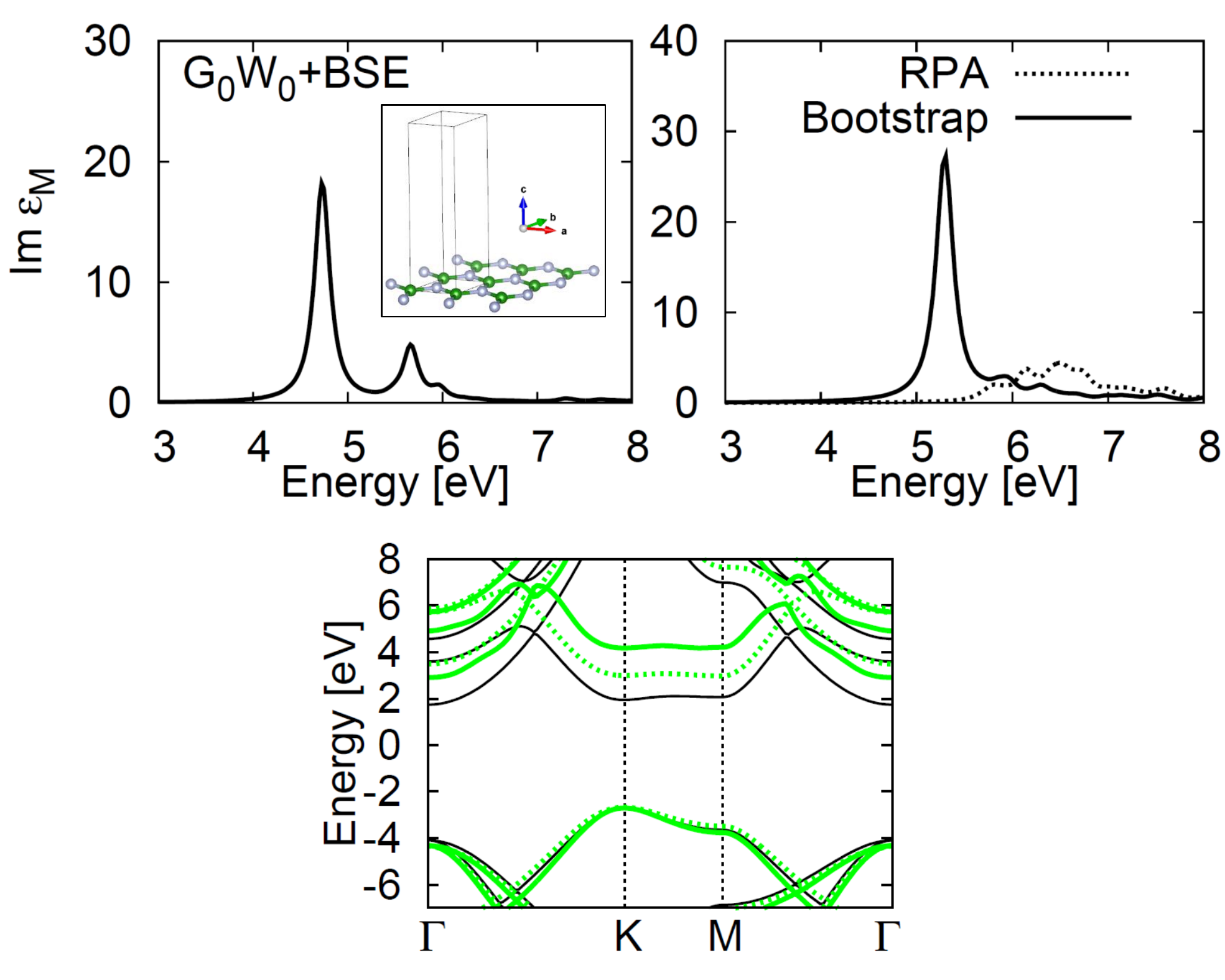}
\end{center}
\caption{\label{Fig3}
Absorption spectra for mono-layer h-BN
calculated by the MBPT approach (upper left panel) and the TDDFT
approach (upper right panel, solid line). 
The first peak in the TDDFT spectra are blue-shifted from that in the BSE spectra by about $0.6$ eV.
The RPA result is also shown as a
dotted line in the upper right panel.
The unit cell is shown in the inset of the upper left panel.
The lower panel shows the band structures (black: using the PBE potential,
green dotted: using the TB-mBJ potential, and green solid: using the G$_0$W$_0$ correction to PBE).  
}
\end{figure}

The upper left panel of Fig.~\ref{Fig3} shows the absorption spectrum of mono-layer h-BN calculated by the MBPT approach (G$_0$W$_0$+BSE)
together with the unit cell employed (inset). The length of the vacuum region is again set as $18$ bohr.
The spectrum shows two characteristic peaks.
These structures are in good agreement with those calculated in previous studies~\cite{bse-m-hBN0,bse-m-hBN1,bse-m-hBN2}.
The peaks are red-shifted by about $0.6-0.8$ eV compared to those from previous reports~\cite{bse-m-hBN1,bse-m-hBN2}, which could be attributed to the 
small vacuum region used in the present calculation
and the fact that here we did not use the 2D Coulomb truncation method~\cite{truncation}.

The spectrum calculated by the TDDFT approach (equipped with the Bootstrap XC kernel and TB-mBJ XC potential)
is plotted as a solid line in the upper right panel of Fig.~\ref{Fig3}.
The result of RPA calculation (i.e., setting $f^{\rm ex}_{\rm xc}=0$ in the TDDFT equations)
is also plotted as a dotted line in the upper right panel.
 A comparison of these spectra indicates that
the spectrum calculated by the TDDFT approach 
captures the qualitative features of that from the MBPT calculation,
i.e., the TDDFT spectrum also shows the intense first peak and the relatively weak second peak, which are completely missed in the RPA result.
(We should note that the first peak in the TDDFT spectrum is blue-shifted by about $0.6$ eV from that in the MBPT spectrum.)

To take a closer look at the origin of this possible agreement  between the MBPT and TDDFT spectra,
we plot the band structures used as a starting point for each calculation in the lower panel of Fig.~\ref{Fig3}
(black: using the PBE potential,
green dotted: using the TB-mBJ potential, and green solid: using the G$_0$W$_0$ correction to PBE).
From this figure, it is seen that the TB-mBJ potential indeed opens up the band gap (at the K point) by about $1.0$ eV 
compared to that calculated using the PBE potential,
while the G$_0$W$_0$ correction opens up the gap by  about $2.0$ eV. 
Thus, the TB-mBJ potential indeed improves the band gap, but it still underestimates the correct value by about $1.0$ eV. 
This implies that the Bootstrap kernel underestimates the exciton binding energy by about $1.6$ eV.
Nevertheless, the present strategy to model $\chi_{\rm qp}$ using the TB-mBJ potential and incorporate the excitonic effect through the use of the Bootstrap kernel is found to work to some extent and
significantly improves the RPA results,  
 and is considered to be effective to investigate the characteristics of the excitonic peaks in the absorption spectrum of 2D mono-layer h-BN.

Having demonstrated that TDDFT approach can extract the overall features of the absorption spectrum of mono-layer h-BN, 
we now present the results for the bi-layer h-BNs.
\begin{figure}[h]
\begin{center}
   \includegraphics[width=1.0\columnwidth]{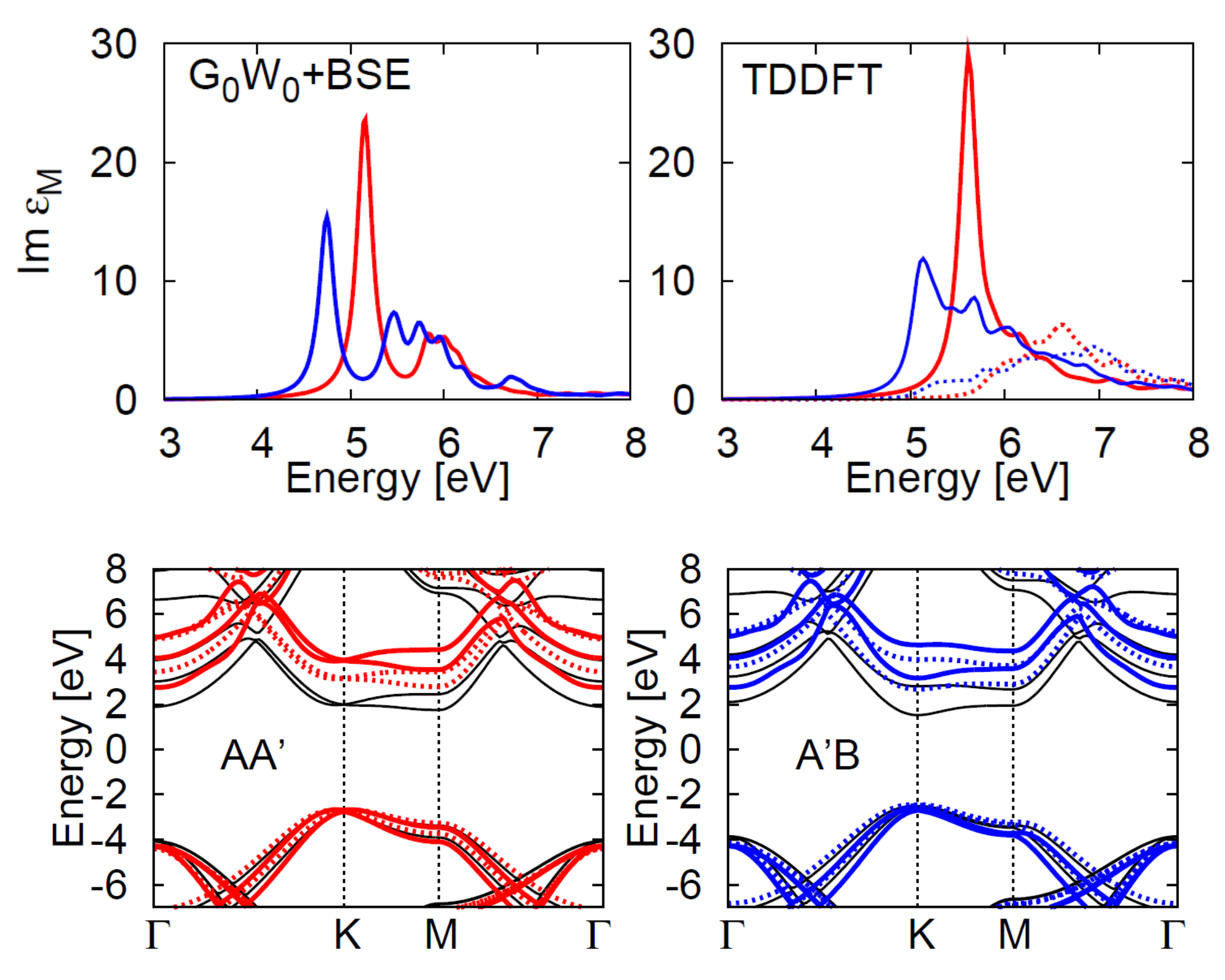}
\end{center}
\caption{\label{Fig4}
Absorption spectra for two different layer-stacking bi-layer h-BNs
calculated by the MBPT approach (upper left panel, the same as Fig.~\ref{Fig2}(c)) and the TDDFT
approach (upper right panel, solid lines). Red lines are for AA' stacking and blue lines are for A'B stacking. 
The first peak in the TDDFT spectra is blue-shifted from that in the BSE spectra by $0.45$ eV for AA' stacking and by $0.40$ eV for A'B stacking.
The RPA results are also shown as a 
dotted line in the upper right panel. 
The lower panel (left: for AA' stacking, right: for A'B stacking)) shows the band structures (black: using the PBE potential,
red or blue dotted: using the TB-mBJ potential, and red or blue solid: using the G$_0$W$_0$ correction to PBE).  
}
\end{figure}
The optical absorption spectra for the AA' (red line) and A'B (blue line) stacking bi-layer h-BNs calculated by the TDDFT approach 
are shown as solid lines in the upper right panel of Fig.~\ref{Fig4}.
For comparison, the MBPT results (Fig.~\ref{Fig2} (c)) are again shown in the upper left panel.
The RPA results are also plotted as dotted lines in the upper right panel.
The lower left and right panel shows the band structures for AA' and A'B stacking (black: PBE,
red or blue dotted: TB-mBJ, and red or blue solid: G$_0$W$_0$ correction to PBE).
From a comparison of the absorption spectra, it is evident that 
the TDDFT approach with the Bootstrap kernel does qualitatively well describe the difference in the
excitonic peaks between the AA' and A'B stacking arrangements,
i.e., the difference between the position and intensity of the first peak for AA' stacking (which corresponds to the intralayer exciton) and those of the first peak for A'B stacking (which corresponds to the
interlayer exciton), 
although the TDDFT peaks are again blue-shifted from the MBPT peaks.  
More specifically, the shift of the first peak is $0.45$ eV for AA' stacking and $0.40$ eV for A'B stacking.
From a comparison of band structures in the lower panels, it is again understood that
the TB-mBJ potential improves the band gap both for AA' and A'B stacking by about $1.2$ eV and $1.25$ eV respectively (at the K point), but still underestimates the correct values
by $0.8$ eV and $0.5$ eV respectively, and 
the Bootsrtap kernel underestimates the exciton binding energies by about $1.25$ eV and $0.9$ eV respectively.
Nevertheless, it can be concluded that the TDDFT approach again significantly improves the RPA results, 
as the RPA calculation does not describe the peak structures correctly. 

These results indicate that the TDDFT approach with the Bootstrap kernel and TB-mBJ potential are
useful to simulate the difference in the absorption spectra between the different layer-stacking bi-layer h-BNs.
 The calculated spectra can be used for qualitative discussions, for example, prediction of the difference in the 
onset of absorption.
It is also demonstrated that the TDDFT approach can capture the excitonic peaks that correspond to the interlayer exciton, which is important 
when considering the application to photovoltaics, for example.
The TDDFT approach we employed directly solves the Dyson-like equation (Eq.~(\ref{eqn:dyson})) to obtain the two-point response function,
which provides the excitonic peak energies.
If the TDDFT calculation is performed via the Casida-type formulation~\cite{fxc11}, one may obtain the eigenvectors that correspond to the excitonic peak energies and use it for the expansion in electron-hole pairs to visualize how the TDDFT approach captures the nature of excitons.
It would be a computationally challenging task for periodic systems
but an important future direction.
We note that the differences between the TDDFT and MBPT spectra might arise from the fact that
our TDDFT approach lacks the non-local exchange kernel~\cite{EXX} that might be necessary for the accurate description of charge transfer, 
as mentioned in the {\bf Formalism section} above.
We also note that 
in pure 2D systems the Coulomb potential and limiting constraint 
for the XC kernel should not be $\sim1/q^2$.
One direction to improve the XC kernel for pure 2D systems is 
to modify the Bootstrap kernel such that it satisfies the proper limiting constraint for 2D Coulomb potential.

Finally, we present the results of TDDFT calculations for other 2D materials, i.e., mono-layer 
MoS$_2$ and mono-layer MoSe$_2$, to show the
usefulness of the TDDFT approach to the study of the optical properties of general 2D materials.
The few-layer TMDCs have been considerably remarked as post-graphene materials
and the building blocks of vdW 
heterostructures~\cite{trion1,trion2,trion3,tmdc1,tmdc2,tmdc3,optel1,optel2,optel3,harvest1,harvest2,harvest3,harvest4,detect1,detect2,emit1,
emit2,emit3,opt1,opt2,opt3,opt4,opt5,opt6,opt8,opt10,dope1,dope2,dope3,tmdcexp}, 
and their application to nano-optoelectronic devices is highly expected.
\begin{figure}[h]
 \centering
 \includegraphics*[width=1.0\columnwidth]{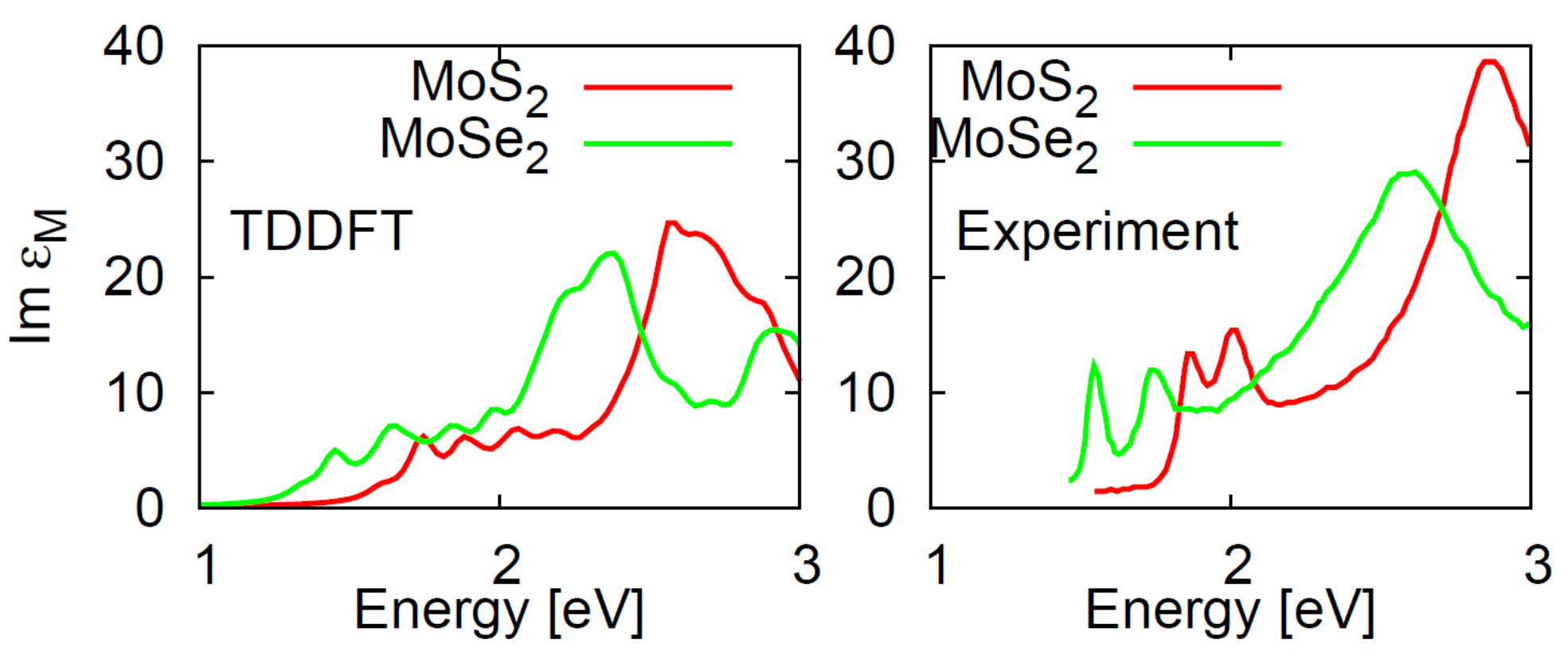}
 \caption{Absorption spectra of mono-layer MoS$_2$ (red) and mono-layer MoSe$_2$ (green) calculated by
the TDDFT (LDA potential + Bootstrap kernel) approach (left panel).
The right panel shows the experimental results~\cite{tmdcexp}.
The TDDFT peaks are red-shifted by about $0.2$ eV from the experimental peaks both for MoS$_2$ and MoSe$_2$.
The TDDFT peaks would be blue-shifted by about 0.8 eV compared to the experimental ones
if the quasiparticle effect were incorporated according to the quasiparticle band structures reported in reference~\cite{bse0}.
}
 \label{Fig5}
\end{figure}
The absorption spectra of mono-layer MoS$_2$ (red) and mono-layer MoSe$_2$ (green) calculated by
the TDDFT approach are shown in the left panel of Fig.~\ref{Fig5}. 
It should be noted that the spin-orbit coupling is explicitly taken into account in these calculations because the heavy element molybdenum is included in the system.
As a result the TB-mBJ potential functional cannot be used~\cite{tb1,tb2,elk}; therefore, the LDA XC potential~\cite{lda} is used for these calculations and only the effect of the Bootstrap kernel is studied. 
The right panel of Fig.~\ref{Fig5} shows the experimental results~\cite{tmdcexp}.
A comparison of the left and right panels indicates that the TDDFT approach again  
captures the qualitative features of the absorption peaks,
although the TDDFT peaks are red-shifted by about $0.2$ eV from the experimental peaks both for MoS$_2$ and MoSe$_2$.
The reason for these red-shifts of TDDFT peaks is considered to be the use of the LDA potential, i.e., 
the lack of the quasiparticle effect. 
However, the peaks would be blue-shifted by about 1.0 eV if the quasiparticle effect were incorporated by the GW method 
(according to the quasiparticle band structures of MoS$_2$ and MoSe$_2$ reported in reference~\cite{bse0}).
Therefore, the TDDFT peaks would be blue-shifted by about 0.8 eV compared to the experimental ones after the
quasiparticle correction. It indicates that our TDDFT approach with the Bootstrap kernel again underestimates the
exciton binding energy, which 
might be attributed to the shortcomings of the Bootstrap kernel discussed above.
The excessive splitting of the excitonic peaks in the TDDFT spectra (left panel of Fig.~\ref{Fig5}) is attributed to the spin-orbit coupling, which can be eliminated by
taking account of the lifetime effect due to electron-phonon coupling~\cite{bse1}.
We note that in the TDDFT results (left panel of Fig.~\ref{Fig5}), the non-Rydberg excitonic series~\cite{rydberg1,rydberg2}, the peaks of which were reported to be
very weak, cannot be recognized. 
The study on whether the refined effects such as the non-Rydberg excitonic series and other interesting exciton
behaviors such as trions~\cite{trion1,trion2,trion3}
can be caught by the TDDFT approach is also an important direction
for future work.

\section{Summary} \label{s:summary}

We have applied the TDDFT approach with the TB-mBJ potential functional and the Bootstrap kernel
to the calculation of absorption spectra for 2D atomic layer materials, mono-layer h-BN, bi-layer h-BNs, and mono-layer MoS$_2$ and MoSe$_2$.
The TDDFT approach describes the overall features of the excitonic peaks of these materials 
with moderate computational cost
although it underestimates the exciton binding energy compared to the MBPT approach.
In particular, for the bi-layer h-BNs, the TDDFT approach can describe the excitonic peaks that correspond to the interlayer exciton and intralayer exciton, and captures the difference in the absorption spectra between the different layer stacking arrangements.
These results
indicate that the TDDFT approach with the Bootstrap kernel and TB-mBJ potential
partially captures the features of the exact XC kernel (described in the {\bf Formalism section} above)
that are essential to describe the quasiparticle and excitonic effects in these 2D materials.

This study demonstrates the usefulness of the TDDFT approach for the qualitative study of the optical properties of 2D materials.
The TDDFT approach has advantages over the MBPT approach in terms of computational efficiency.
Improving the XC terms will improve the accuracy of the TDDFT approach, so that it will become one of the key simulation tools in the 
development of 2D material science.

\section*{Conflicts of interest}
There are no conflicts to declare.

\section*{Acknowledgements}
Y. S. is supported by JSPS KAKENHI Grant No. JP16K17768 and JP19K03675.
K. W. is supported by JSPS KAKENHI Grant No. JP16K05483.
Part of the computations were performed on
the supercomputers of the Institute for Solid State Physics,
The University of Tokyo.



\balance


\bibliography{reference} 

\providecommand*{\mcitethebibliography}{\thebibliography}
\csname @ifundefined\endcsname{endmcitethebibliography}
{\let\endmcitethebibliography\endthebibliography}{}
\begin{mcitethebibliography}{117}
\providecommand*{\natexlab}[1]{#1}
\providecommand*{\mciteSetBstSublistMode}[1]{}
\providecommand*{\mciteSetBstMaxWidthForm}[2]{}
\providecommand*{\mciteBstWouldAddEndPuncttrue}
  {\def\EndOfBibitem{\unskip.}}
\providecommand*{\mciteBstWouldAddEndPunctfalse}
  {\let\EndOfBibitem\relax}
\providecommand*{\mciteSetBstMidEndSepPunct}[3]{}
\providecommand*{\mciteSetBstSublistLabelBeginEnd}[3]{}
\providecommand*{\EndOfBibitem}{}
\mciteSetBstSublistMode{f}
\mciteSetBstMaxWidthForm{subitem}
{(\emph{\alph{mcitesubitemcount}})}
\mciteSetBstSublistLabelBeginEnd{\mcitemaxwidthsubitemform\space}
{\relax}{\relax}

\bibitem[Novoselov \emph{et~al.}(2004)Novoselov, Geim, Morozov, Jiang, Zhang,
  Dubonos, Grigorieva, and Firsov]{graphene}
K.~S. Novoselov, A.~K. Geim, S.~V. Morozov, D.~Jiang, Y.~Zhang, S.~V. Dubonos,
  I.~V. Grigorieva and A.~A. Firsov, \emph{Science}, 2004, \textbf{306},
  666\relax
\mciteBstWouldAddEndPuncttrue
\mciteSetBstMidEndSepPunct{\mcitedefaultmidpunct}
{\mcitedefaultendpunct}{\mcitedefaultseppunct}\relax
\EndOfBibitem
\bibitem[Novoselov \emph{et~al.}(2005)Novoselov, Geim, Morozov, Jiang,
  Katsnelson, Grigorieva, Dubonos, and Firsov]{hall1}
K.~S. Novoselov, A.~K. Geim, S.~V. Morozov, D.~Jiang, M.~I. Katsnelson, I.~V.
  Grigorieva, S.~V. Dubonos and A.~A. Firsov, \emph{Nature}, 2005,
  \textbf{438}, 197\relax
\mciteBstWouldAddEndPuncttrue
\mciteSetBstMidEndSepPunct{\mcitedefaultmidpunct}
{\mcitedefaultendpunct}{\mcitedefaultseppunct}\relax
\EndOfBibitem
\bibitem[Zhang \emph{et~al.}(2005)Zhang, Tan, Stormer, and Kim]{hall2}
Y.~Zhang, Y.-W. Tan, H.~L. Stormer and P.~Kim, \emph{Nature}, 2005,
  \textbf{438}, 201\relax
\mciteBstWouldAddEndPuncttrue
\mciteSetBstMidEndSepPunct{\mcitedefaultmidpunct}
{\mcitedefaultendpunct}{\mcitedefaultseppunct}\relax
\EndOfBibitem
\bibitem[Mak \emph{et~al.}(2013)Mak, He, Lee, Lee, Hone, Heinz, and
  Shan]{trion1}
K.~F. Mak, K.~He, C.~Lee, G.~H. Lee, J.~Hone, T.~F. Heinz and J.~Shan,
  \emph{Nat. Mater.}, 2013, \textbf{12}, 207\relax
\mciteBstWouldAddEndPuncttrue
\mciteSetBstMidEndSepPunct{\mcitedefaultmidpunct}
{\mcitedefaultendpunct}{\mcitedefaultseppunct}\relax
\EndOfBibitem
\bibitem[Plechinger \emph{et~al.}(2016)Plechinger, Nagler, Arora, Schmidt,
  Chernikov, {A. G. del {\'A}guila}, Christianen, Bratschitsch, Sch{\"u}ller,
  and Korn]{trion2}
G.~Plechinger, P.~Nagler, A.~Arora, R.~Schmidt, A.~Chernikov, {A. G. del
  {\'A}guila}, P.~C.~M. Christianen, R.~Bratschitsch, C.~Sch{\"u}ller and
  T.~Korn, \emph{Nat. Commun.}, 2016, \textbf{7}, 12715\relax
\mciteBstWouldAddEndPuncttrue
\mciteSetBstMidEndSepPunct{\mcitedefaultmidpunct}
{\mcitedefaultendpunct}{\mcitedefaultseppunct}\relax
\EndOfBibitem
\bibitem[Ramirez-Torres \emph{et~al.}(2014)Ramirez-Torres, Turkowski, and
  Rahman]{trion3}
A.~Ramirez-Torres, V.~Turkowski and T.~S. Rahman, \emph{Phys. Rev. B}, 2014,
  \textbf{90}, 085419\relax
\mciteBstWouldAddEndPuncttrue
\mciteSetBstMidEndSepPunct{\mcitedefaultmidpunct}
{\mcitedefaultendpunct}{\mcitedefaultseppunct}\relax
\EndOfBibitem
\bibitem[Behnia(2012)]{valley1}
K.~Behnia, \emph{Nat. Nanotechnol.}, 2012, \textbf{7}, 488\relax
\mciteBstWouldAddEndPuncttrue
\mciteSetBstMidEndSepPunct{\mcitedefaultmidpunct}
{\mcitedefaultendpunct}{\mcitedefaultseppunct}\relax
\EndOfBibitem
\bibitem[Schaibley \emph{et~al.}(2016)Schaibley, Yu, Clark, Rivera, Ross,
  Seyler, Yao, and Xu]{valley2}
J.~R. Schaibley, H.~Yu, G.~Clark, P.~Rivera, J.~S. Ross, K.~L. Seyler, W.~Yao
  and X.~Xu, \emph{Nat. Rev. Mater.}, 2016, \textbf{1}, 16055\relax
\mciteBstWouldAddEndPuncttrue
\mciteSetBstMidEndSepPunct{\mcitedefaultmidpunct}
{\mcitedefaultendpunct}{\mcitedefaultseppunct}\relax
\EndOfBibitem
\bibitem[Miyauchi \emph{et~al.}(2018)Miyauchi, Konabe, Wang, Zhang, Hwang,
  Hasegawa, Zhou, Mouri, Toh, Eda, and Matsuda]{valley3}
Y.~Miyauchi, S.~Konabe, F.~Wang, W.~Zhang, A.~Hwang, Y.~Hasegawa, L.~Zhou,
  S.~Mouri, M.~Toh, G.~Eda and K.~Matsuda, \emph{Nat. Commun.}, 2018,
  \textbf{9}, 2598\relax
\mciteBstWouldAddEndPuncttrue
\mciteSetBstMidEndSepPunct{\mcitedefaultmidpunct}
{\mcitedefaultendpunct}{\mcitedefaultseppunct}\relax
\EndOfBibitem
\bibitem[Splendiani \emph{et~al.}(2010)Splendiani, Sun, Zhang, Li, Kim, Chim,
  Galli, and Wang]{tmdc1}
A.~Splendiani, L.~Sun, Y.~Zhang, T.~Li, J.~Kim, C.-Y. Chim, G.~Galli and
  F.~Wang, \emph{Nano Lett.}, 2010, \textbf{10}, 1271\relax
\mciteBstWouldAddEndPuncttrue
\mciteSetBstMidEndSepPunct{\mcitedefaultmidpunct}
{\mcitedefaultendpunct}{\mcitedefaultseppunct}\relax
\EndOfBibitem
\bibitem[Radisavljevic \emph{et~al.}(2011)Radisavljevic, Radenovi, Brivio,
  Giacometti, and Kis]{tmdc2}
B.~Radisavljevic, A.~Radenovi, J.~Brivio, V.~Giacometti and A.~Kis, \emph{Nat.
  Nanotechnol.}, 2011, \textbf{6}, 147\relax
\mciteBstWouldAddEndPuncttrue
\mciteSetBstMidEndSepPunct{\mcitedefaultmidpunct}
{\mcitedefaultendpunct}{\mcitedefaultseppunct}\relax
\EndOfBibitem
\bibitem[Chhowalla \emph{et~al.}(2013)Chhowalla\emph{et~al.}]{tmdc3}
M.~Chhowalla \emph{et~al.}, \emph{Nat. Chem.}, 2013, \textbf{5}, 263\relax
\mciteBstWouldAddEndPuncttrue
\mciteSetBstMidEndSepPunct{\mcitedefaultmidpunct}
{\mcitedefaultendpunct}{\mcitedefaultseppunct}\relax
\EndOfBibitem
\bibitem[Stehle \emph{et~al.}(2015)Stehle, Meyer~III, Unocic, Kidder, Polizos,
  Datskos, Jackson, Smirnov, and Vlassiouk]{hBN1}
Y.~Stehle, H.~M. Meyer~III, R.~R. Unocic, M.~Kidder, G.~Polizos, P.~G. Datskos,
  R.~Jackson, S.~N. Smirnov and I.~V. Vlassiouk, \emph{Chem. Mater.}, 2015,
  \textbf{27}, 8041\relax
\mciteBstWouldAddEndPuncttrue
\mciteSetBstMidEndSepPunct{\mcitedefaultmidpunct}
{\mcitedefaultendpunct}{\mcitedefaultseppunct}\relax
\EndOfBibitem
\bibitem[Wu \emph{et~al.}(2014)Wu, Schmidt, Amara, Xu, Eda, and
  {\"O}zyilmaz]{thermo0}
J.~Wu, H.~Schmidt, K.~K. Amara, X.~Xu, G.~Eda and B.~{\"O}zyilmaz, \emph{Nano
  Lett.}, 2014, \textbf{14}, 2730\relax
\mciteBstWouldAddEndPuncttrue
\mciteSetBstMidEndSepPunct{\mcitedefaultmidpunct}
{\mcitedefaultendpunct}{\mcitedefaultseppunct}\relax
\EndOfBibitem
\bibitem[Chen \emph{et~al.}(2017)Chen, Lyu, Wang, Fu, Heng, and Mo]{thermo1}
K.-X. Chen, S.-S. Lyu, X.-M. Wang, Y.-X. Fu, Y.~Heng and D.-C. Mo, \emph{J.
  Phys. Chem. C.}, 2017, \textbf{121}, 13035\relax
\mciteBstWouldAddEndPuncttrue
\mciteSetBstMidEndSepPunct{\mcitedefaultmidpunct}
{\mcitedefaultendpunct}{\mcitedefaultseppunct}\relax
\EndOfBibitem
\bibitem[{Ch. Adessi} \emph{et~al.}(2017){Ch. Adessi}, Thebaud, Bouzerar, and
  Bouzerar]{thermo2}
{Ch. Adessi}, S.~Thebaud, R.~Bouzerar and G.~Bouzerar, \emph{J. Phys. Chem.
  C.}, 2017, \textbf{121}, 12577\relax
\mciteBstWouldAddEndPuncttrue
\mciteSetBstMidEndSepPunct{\mcitedefaultmidpunct}
{\mcitedefaultendpunct}{\mcitedefaultseppunct}\relax
\EndOfBibitem
\bibitem[Zhu \emph{et~al.}(2017)Zhu, Zou, Gao, and Yao]{thermo3}
L.~Zhu, F.~Zou, G.~Gao and K.~Yao, \emph{Sci. Rep.}, 2017, \textbf{7},
  497\relax
\mciteBstWouldAddEndPuncttrue
\mciteSetBstMidEndSepPunct{\mcitedefaultmidpunct}
{\mcitedefaultendpunct}{\mcitedefaultseppunct}\relax
\EndOfBibitem
\bibitem[Wang \emph{et~al.}(2012)Wang, Kalantar-Zadeh, Kis, Coleman, and
  Strano]{optel1}
Q.~H. Wang, K.~Kalantar-Zadeh, A.~Kis, J.~N. Coleman and M.~S. Strano,
  \emph{Nat. Nanotechnol.}, 2012, \textbf{7}, 699\relax
\mciteBstWouldAddEndPuncttrue
\mciteSetBstMidEndSepPunct{\mcitedefaultmidpunct}
{\mcitedefaultendpunct}{\mcitedefaultseppunct}\relax
\EndOfBibitem
\bibitem[Mak and Shan(2016)]{optel2}
K.~F. Mak and J.~Shan, \emph{Nat. Photonics}, 2016, \textbf{10}, 216\relax
\mciteBstWouldAddEndPuncttrue
\mciteSetBstMidEndSepPunct{\mcitedefaultmidpunct}
{\mcitedefaultendpunct}{\mcitedefaultseppunct}\relax
\EndOfBibitem
\bibitem[Jariwala \emph{et~al.}(2014)Jariwala, Sangwan, Lauhon, Marks, and
  Hersam]{optel3}
D.~Jariwala, V.~K. Sangwan, L.~J. Lauhon, T.~J. Marks and M.~C. Hersam,
  \emph{ACS Nano}, 2014, \textbf{8}, 1102\relax
\mciteBstWouldAddEndPuncttrue
\mciteSetBstMidEndSepPunct{\mcitedefaultmidpunct}
{\mcitedefaultendpunct}{\mcitedefaultseppunct}\relax
\EndOfBibitem
\bibitem[Tsai \emph{et~al.}(2014)Tsai, Su, Chang, Tsai, Chen, Wu, Li, Chen, and
  He]{harvest1}
M.-L. Tsai, S.-H. Su, J.-K. Chang, D.-S. Tsai, C.-H. Chen, C.-I. Wu, L.-J. Li,
  L.-J. Chen and J.-H. He, \emph{ACS Nano}, 2014, \textbf{8}, 8317\relax
\mciteBstWouldAddEndPuncttrue
\mciteSetBstMidEndSepPunct{\mcitedefaultmidpunct}
{\mcitedefaultendpunct}{\mcitedefaultseppunct}\relax
\EndOfBibitem
\bibitem[Bernardi \emph{et~al.}(2013)Bernardi, Palummo, and Grossman]{harvest2}
M.~Bernardi, M.~Palummo and J.~C. Grossman, \emph{Nano Lett.}, 2013,
  \textbf{13}, 3664\relax
\mciteBstWouldAddEndPuncttrue
\mciteSetBstMidEndSepPunct{\mcitedefaultmidpunct}
{\mcitedefaultendpunct}{\mcitedefaultseppunct}\relax
\EndOfBibitem
\bibitem[Britnell \emph{et~al.}(2013)Britnell, Bibeiro, Eckmann, Jalil, Belle,
  Mishchenko, Kim, Gorbachev, Georgiou, Morozov, Grigorenko, Geim, Casiraghi,
  Castro~Neto, and Novoselov]{harvest3}
L.~Britnell, R.~M. Bibeiro, A.~Eckmann, R.~Jalil, B.~D. Belle, A.~Mishchenko,
  Y.-J. Kim, R.~V. Gorbachev, T.~Georgiou, S.~V. Morozov, A.~N. Grigorenko,
  A.~K. Geim, C.~Casiraghi, A.~H. Castro~Neto and K.~S. Novoselov,
  \emph{Science}, 2013, \textbf{340}, 1311\relax
\mciteBstWouldAddEndPuncttrue
\mciteSetBstMidEndSepPunct{\mcitedefaultmidpunct}
{\mcitedefaultendpunct}{\mcitedefaultseppunct}\relax
\EndOfBibitem
\bibitem[Lopez-Sanchez \emph{et~al.}(2014)Lopez-Sanchez, Alarcon~Llado, Koman,
  Fontcubertai~Morral, Radenovic, and Kis]{harvest4}
O.~Lopez-Sanchez, E.~Alarcon~Llado, V.~Koman, A.~Fontcubertai~Morral,
  A.~Radenovic and A.~Kis, \emph{ACS Nano}, 2014, \textbf{8}, 3042\relax
\mciteBstWouldAddEndPuncttrue
\mciteSetBstMidEndSepPunct{\mcitedefaultmidpunct}
{\mcitedefaultendpunct}{\mcitedefaultseppunct}\relax
\EndOfBibitem
\bibitem[Lee \emph{et~al.}(2012)Lee, Min, Chang, Park, Nam, Kim, Kim, Ryu, and
  Im]{detect1}
H.~S. Lee, S.~W. Min, Y.~G. Chang, M.~K. Park, T.~Nam, H.~Kim, J.~H. Kim,
  S.~Ryu and S.~Im, \emph{Nano Lett.}, 2012, \textbf{12}, 3695\relax
\mciteBstWouldAddEndPuncttrue
\mciteSetBstMidEndSepPunct{\mcitedefaultmidpunct}
{\mcitedefaultendpunct}{\mcitedefaultseppunct}\relax
\EndOfBibitem
\bibitem[Xia \emph{et~al.}(2014)Xia, Huang, Liu, Wang, Wang, Huang, Zhu, J.-J,
  Gu, and Meng]{detect2}
J.~Xia, X.~Huang, L.-Z. Liu, M.~Wang, L.~Wang, B.~Huang, D.-D. Zhu, L.~J.-J,
  C.-Z. Gu and X.-M. Meng, \emph{Nanoscale}, 2014, \textbf{6}, 8949\relax
\mciteBstWouldAddEndPuncttrue
\mciteSetBstMidEndSepPunct{\mcitedefaultmidpunct}
{\mcitedefaultendpunct}{\mcitedefaultseppunct}\relax
\EndOfBibitem
\bibitem[Sundaram \emph{et~al.}(2013)Sundaram, Engel, Lombardo, Krupke,
  Ferrari, {Ph. Avouris}, and Steiner]{emit1}
R.~S. Sundaram, M.~Engel, A.~Lombardo, R.~Krupke, A.~C. Ferrari, {Ph. Avouris}
  and M.~Steiner, \emph{Nano Lett.}, 2013, \textbf{13}, 1416\relax
\mciteBstWouldAddEndPuncttrue
\mciteSetBstMidEndSepPunct{\mcitedefaultmidpunct}
{\mcitedefaultendpunct}{\mcitedefaultseppunct}\relax
\EndOfBibitem
\bibitem[Zhang \emph{et~al.}(2014)Zhang, Oka, Suzuki, Ye, and Iwasa]{emit2}
Y.~J. Zhang, T.~Oka, R.~Suzuki, J.~T. Ye and Y.~Iwasa, \emph{Science}, 2014,
  \textbf{344}, 725\relax
\mciteBstWouldAddEndPuncttrue
\mciteSetBstMidEndSepPunct{\mcitedefaultmidpunct}
{\mcitedefaultendpunct}{\mcitedefaultseppunct}\relax
\EndOfBibitem
\bibitem[Ross \emph{et~al.}(2014)Ross, Klement, Jones, Ghimire, Yan, Mandrus,
  Taniguchi, Watanabe, Kitamura, Yao, Cobden, and Xu]{emit3}
J.~S. Ross, P.~Klement, A.~M. Jones, N.~J. Ghimire, J.~Yan, D.~Mandrus,
  T.~Taniguchi, K.~Watanabe, K.~Kitamura, W.~Yao, D.~H. Cobden and X.~Xu,
  \emph{Nat. Nanotechnol.}, 2014, \textbf{9}, 268\relax
\mciteBstWouldAddEndPuncttrue
\mciteSetBstMidEndSepPunct{\mcitedefaultmidpunct}
{\mcitedefaultendpunct}{\mcitedefaultseppunct}\relax
\EndOfBibitem
\bibitem[Bourrellier \emph{et~al.}(2014)Bourrellier, Amato, Henrique, Tizei,
  Giorgetti, Gloter, Heggie, March, St{\'e}phan, Reining, Kociak, and
  Zobelli]{emit4}
R.~Bourrellier, M.~Amato, L.~Henrique, G.~Tizei, C.~Giorgetti, A.~Gloter, M.~I.
  Heggie, K.~March, O.~St{\'e}phan, L.~Reining, M.~Kociak and A.~Zobelli,
  \emph{ACS Photonics}, 2014, \textbf{1}, 857\relax
\mciteBstWouldAddEndPuncttrue
\mciteSetBstMidEndSepPunct{\mcitedefaultmidpunct}
{\mcitedefaultendpunct}{\mcitedefaultseppunct}\relax
\EndOfBibitem
\bibitem[Geim and Grigorieva(2013)]{vdW1}
A.~K. Geim and I.~V. Grigorieva, \emph{Nature}, 2013, \textbf{499}, 419\relax
\mciteBstWouldAddEndPuncttrue
\mciteSetBstMidEndSepPunct{\mcitedefaultmidpunct}
{\mcitedefaultendpunct}{\mcitedefaultseppunct}\relax
\EndOfBibitem
\bibitem[Wang \emph{et~al.}(2017)Wang, Ma, and Sun]{vdW2}
J.~Wang, F.~Ma and M.~Sun, \emph{RSC Adv.}, 2017, \textbf{7}, 16801\relax
\mciteBstWouldAddEndPuncttrue
\mciteSetBstMidEndSepPunct{\mcitedefaultmidpunct}
{\mcitedefaultendpunct}{\mcitedefaultseppunct}\relax
\EndOfBibitem
\bibitem[Wang \emph{et~al.}(2017)Wang, Ma, Liang, and Sun]{vdW3}
J.~Wang, F.~Ma, W.~Liang and M.~Sun, \emph{Materials Today Physics}, 2017,
  \textbf{2}, 6\relax
\mciteBstWouldAddEndPuncttrue
\mciteSetBstMidEndSepPunct{\mcitedefaultmidpunct}
{\mcitedefaultendpunct}{\mcitedefaultseppunct}\relax
\EndOfBibitem
\bibitem[Wang \emph{et~al.}(2017)Wang, Xu, Mu, Ma, and Sun]{vdW4}
J.~Wang, X.~Xu, X.~Mu, F.~Ma and M.~Sun, \emph{Materials Today Physics}, 2017,
  \textbf{3}, 93\relax
\mciteBstWouldAddEndPuncttrue
\mciteSetBstMidEndSepPunct{\mcitedefaultmidpunct}
{\mcitedefaultendpunct}{\mcitedefaultseppunct}\relax
\EndOfBibitem
\bibitem[Mu \emph{et~al.}(2019)Mu, Wang, and Sun]{vdW5}
X.~Mu, J.~Wang and M.~Sun, \emph{Materials Today Physics}, 2019, \textbf{8},
  92\relax
\mciteBstWouldAddEndPuncttrue
\mciteSetBstMidEndSepPunct{\mcitedefaultmidpunct}
{\mcitedefaultendpunct}{\mcitedefaultseppunct}\relax
\EndOfBibitem
\bibitem[Kozawa \emph{et~al.}(2014)Kozawa, Kumar, Carvalho, Amara, Zhao, Wang,
  Toh, Rbeiro, Castro~Neto, Matsuda, and Eda]{opt1}
D.~Kozawa, R.~Kumar, A.~Carvalho, K.~K. Amara, W.~Zhao, S.~Wang, M.~Toh, R.~M.
  Rbeiro, A.~H. Castro~Neto, K.~Matsuda and G.~Eda, \emph{Nat. Commun.}, 2014,
  \textbf{5}, 4543\relax
\mciteBstWouldAddEndPuncttrue
\mciteSetBstMidEndSepPunct{\mcitedefaultmidpunct}
{\mcitedefaultendpunct}{\mcitedefaultseppunct}\relax
\EndOfBibitem
\bibitem[Mouri \emph{et~al.}(2017)Mouri, Zhang, Kozawa, Miyauchi, Eda, and
  Matsuda]{opt2}
S.~Mouri, W.~Zhang, D.~Kozawa, Y.~Miyauchi, G.~Eda and K.~Matsuda,
  \emph{Nanoscale}, 2017, \textbf{9}, 6674\relax
\mciteBstWouldAddEndPuncttrue
\mciteSetBstMidEndSepPunct{\mcitedefaultmidpunct}
{\mcitedefaultendpunct}{\mcitedefaultseppunct}\relax
\EndOfBibitem
\bibitem[Novoselov \emph{et~al.}(2016)Novoselov, Mishcenko, Carvalho, and {A.
  H. Castro Neto}]{opt3}
K.~S. Novoselov, A.~Mishcenko, A.~Carvalho and {A. H. Castro Neto},
  \emph{Science}, 2016, \textbf{353}, aac9439\relax
\mciteBstWouldAddEndPuncttrue
\mciteSetBstMidEndSepPunct{\mcitedefaultmidpunct}
{\mcitedefaultendpunct}{\mcitedefaultseppunct}\relax
\EndOfBibitem
\bibitem[Jariwala \emph{et~al.}(2017)Jariwala, Marks, and Hersam]{opt4}
D.~Jariwala, T.~J. Marks and M.~C. Hersam, \emph{Nat. Mater.}, 2017,
  \textbf{16}, 170\relax
\mciteBstWouldAddEndPuncttrue
\mciteSetBstMidEndSepPunct{\mcitedefaultmidpunct}
{\mcitedefaultendpunct}{\mcitedefaultseppunct}\relax
\EndOfBibitem
\bibitem[Okada \emph{et~al.}(2018)Okada, Kutana, Kureishi, Kobayashi, Saito,
  Saito, Watanabe, Taniguchi, Gupta, Miyata, Yakobson, Shinohara, and
  Kitaura]{opt5}
M.~Okada, A.~Kutana, Y.~Kureishi, Y.~Kobayashi, Y.~Saito, T.~Saito,
  K.~Watanabe, T.~Taniguchi, S.~Gupta, Y.~Miyata, B.~I. Yakobson, H.~Shinohara
  and R.~Kitaura, \emph{ACS Nano}, 2018, \textbf{12}, 2498\relax
\mciteBstWouldAddEndPuncttrue
\mciteSetBstMidEndSepPunct{\mcitedefaultmidpunct}
{\mcitedefaultendpunct}{\mcitedefaultseppunct}\relax
\EndOfBibitem
\bibitem[Nourbakhsh \emph{et~al.}(2016)Nourbakhsh, Zubair, Dresselhaus, and
  Palacios]{opt6}
A.~Nourbakhsh, A.~Zubair, M.~S. Dresselhaus and T.~Palacios, \emph{Nano Lett.},
  2016, \textbf{16}, 1359\relax
\mciteBstWouldAddEndPuncttrue
\mciteSetBstMidEndSepPunct{\mcitedefaultmidpunct}
{\mcitedefaultendpunct}{\mcitedefaultseppunct}\relax
\EndOfBibitem
\bibitem[Hu and Hong(2015)]{opt7}
T.~Hu and J.~Hong, \emph{ACS Appl. Mater. Interfaces}, 2015, \textbf{7},
  23489\relax
\mciteBstWouldAddEndPuncttrue
\mciteSetBstMidEndSepPunct{\mcitedefaultmidpunct}
{\mcitedefaultendpunct}{\mcitedefaultseppunct}\relax
\EndOfBibitem
\bibitem[Latini \emph{et~al.}(2017)Latini, Winther, Olsen, and Thygesen]{opt8}
S.~Latini, K.~T. Winther, T.~Olsen and K.~S. Thygesen, \emph{Nano Lett.}, 2017,
  \textbf{17}, 938\relax
\mciteBstWouldAddEndPuncttrue
\mciteSetBstMidEndSepPunct{\mcitedefaultmidpunct}
{\mcitedefaultendpunct}{\mcitedefaultseppunct}\relax
\EndOfBibitem
\bibitem[Chen and Quek(2018)]{opt9}
Y.~Chen and S.~Y. Quek, \emph{2D Mater.}, 2018, \textbf{5}, 045031\relax
\mciteBstWouldAddEndPuncttrue
\mciteSetBstMidEndSepPunct{\mcitedefaultmidpunct}
{\mcitedefaultendpunct}{\mcitedefaultseppunct}\relax
\EndOfBibitem
\bibitem[Nayak \emph{et~al.}(2017)Nayak, Horbatenko, Ahn, Kim, Lee, Ma, Jang,
  Lim, Kim, Ryu, Cheong, Park, and Shin]{opt10}
P.~K. Nayak, Y.~Horbatenko, S.~Ahn, G.~Kim, J.-U. Lee, K.~Y. Ma, A.-R. Jang,
  H.~Lim, D.~Kim, S.~Ryu, H.~Cheong, N.~Park and H.~S. Shin, \emph{ACS Nano},
  2017, \textbf{11}, 4041\relax
\mciteBstWouldAddEndPuncttrue
\mciteSetBstMidEndSepPunct{\mcitedefaultmidpunct}
{\mcitedefaultendpunct}{\mcitedefaultseppunct}\relax
\EndOfBibitem
\bibitem[Mouri \emph{et~al.}(2013)Mouri, Miyauchi, and Matsuda]{dope1}
S.~Mouri, Y.~Miyauchi and K.~Matsuda, \emph{Nano Lett.}, 2013, \textbf{13},
  5944\relax
\mciteBstWouldAddEndPuncttrue
\mciteSetBstMidEndSepPunct{\mcitedefaultmidpunct}
{\mcitedefaultendpunct}{\mcitedefaultseppunct}\relax
\EndOfBibitem
\bibitem[Mouri \emph{et~al.}(2016)Mouri, Miyauchi, and Matsuda]{dope2}
S.~Mouri, Y.~Miyauchi and K.~Matsuda, \emph{Appl. Phys. Express}, 2016,
  \textbf{9}, 055202\relax
\mciteBstWouldAddEndPuncttrue
\mciteSetBstMidEndSepPunct{\mcitedefaultmidpunct}
{\mcitedefaultendpunct}{\mcitedefaultseppunct}\relax
\EndOfBibitem
\bibitem[Amani \emph{et~al.}(2015)Amani, Lien, Kiriya, Xiao, Azcatl, Noh,
  Madhvapathy, Addou, KC, Dubey, Cho, Wallace, Lee, He, Ager~III, Zhang,
  Yablonovitch, and Javey]{dope3}
M.~Amani, D.-H. Lien, D.~Kiriya, J.~Xiao, A.~Azcatl, J.~Noh, S.~R. Madhvapathy,
  R.~Addou, S.~KC, M.~Dubey, K.~Cho, R.~M. Wallace, S.-C. Lee, J.-H. He, J.~W.
  Ager~III, X.~Zhang, E.~Yablonovitch and A.~Javey, \emph{Science}, 2015,
  \textbf{350}, 1065\relax
\mciteBstWouldAddEndPuncttrue
\mciteSetBstMidEndSepPunct{\mcitedefaultmidpunct}
{\mcitedefaultendpunct}{\mcitedefaultseppunct}\relax
\EndOfBibitem
\bibitem[Sahin \emph{et~al.}(2016)Sahin, Torun, Bacaksiz, Horzum, Kang, Senger,
  and Peeters]{sim2d1}
H.~Sahin, E.~Torun, C.~Bacaksiz, S.~Horzum, J.~Kang, R.~T. Senger and F.~M.
  Peeters, \emph{WIREs Comput. Mol. Sci.}, 2016, \textbf{6}, 351\relax
\mciteBstWouldAddEndPuncttrue
\mciteSetBstMidEndSepPunct{\mcitedefaultmidpunct}
{\mcitedefaultendpunct}{\mcitedefaultseppunct}\relax
\EndOfBibitem
\bibitem[Adler(1962)]{rpa1}
S.~L. Adler, \emph{Phys. Rev.}, 1962, \textbf{126}, 413\relax
\mciteBstWouldAddEndPuncttrue
\mciteSetBstMidEndSepPunct{\mcitedefaultmidpunct}
{\mcitedefaultendpunct}{\mcitedefaultseppunct}\relax
\EndOfBibitem
\bibitem[Wiser(1963)]{rpa2}
N.~Wiser, \emph{Phys. Rev.}, 1963, \textbf{129}, 62\relax
\mciteBstWouldAddEndPuncttrue
\mciteSetBstMidEndSepPunct{\mcitedefaultmidpunct}
{\mcitedefaultendpunct}{\mcitedefaultseppunct}\relax
\EndOfBibitem
\bibitem[Onida \emph{et~al.}(2002)Onida, Reining, and Rubio]{mbpt1}
G.~Onida, L.~Reining and A.~Rubio, \emph{Rev. Mod. Phys.}, 2002, \textbf{74},
  601\relax
\mciteBstWouldAddEndPuncttrue
\mciteSetBstMidEndSepPunct{\mcitedefaultmidpunct}
{\mcitedefaultendpunct}{\mcitedefaultseppunct}\relax
\EndOfBibitem
\bibitem[Bechstedt(2015)]{mbpt2}
F.~Bechstedt, \emph{Many-Body Approach to Electronic Excitations}, Springer,
  2015\relax
\mciteBstWouldAddEndPuncttrue
\mciteSetBstMidEndSepPunct{\mcitedefaultmidpunct}
{\mcitedefaultendpunct}{\mcitedefaultseppunct}\relax
\EndOfBibitem
\bibitem[Yang \emph{et~al.}(2009)Yang, Deslippe, Park, Cohen, and Louie]{bse-1}
L.~Yang, J.~Deslippe, C.-H. Park, M.~L. Cohen and S.~G. Louie, \emph{Phys. Rev.
  Lett.}, 2009, \textbf{103}, 186802\relax
\mciteBstWouldAddEndPuncttrue
\mciteSetBstMidEndSepPunct{\mcitedefaultmidpunct}
{\mcitedefaultendpunct}{\mcitedefaultseppunct}\relax
\EndOfBibitem
\bibitem[Ramasubramaniam(2012)]{bse0}
A.~Ramasubramaniam, \emph{Phys. Rev. B}, 2012, \textbf{86}, 115409\relax
\mciteBstWouldAddEndPuncttrue
\mciteSetBstMidEndSepPunct{\mcitedefaultmidpunct}
{\mcitedefaultendpunct}{\mcitedefaultseppunct}\relax
\EndOfBibitem
\bibitem[Qiu \emph{et~al.}(2013)Qiu, {F. H. da Jornada}, and Louie]{bse1}
D.~Y. Qiu, {F. H. da Jornada} and S.~G. Louie, \emph{Phys. Rev. Lett.}, 2013,
  \textbf{111}, 216805\relax
\mciteBstWouldAddEndPuncttrue
\mciteSetBstMidEndSepPunct{\mcitedefaultmidpunct}
{\mcitedefaultendpunct}{\mcitedefaultseppunct}\relax
\EndOfBibitem
\bibitem[Molina-S{\'a}nchez \emph{et~al.}(2013)Molina-S{\'a}nchez, Sangalli,
  Hummer, Marini, and Wirtz]{bse2}
A.~Molina-S{\'a}nchez, D.~Sangalli, K.~Hummer, A.~Marini and L.~Wirtz,
  \emph{Phys. Rev. B}, 2013, \textbf{88}, 045412\relax
\mciteBstWouldAddEndPuncttrue
\mciteSetBstMidEndSepPunct{\mcitedefaultmidpunct}
{\mcitedefaultendpunct}{\mcitedefaultseppunct}\relax
\EndOfBibitem
\bibitem[Shi \emph{et~al.}(2013)Shi, Pan, Zhang, and Yakobson]{bse3}
H.~Shi, H.~Pan, Y.-W. Zhang and B.~I. Yakobson, \emph{Phys. Rev. B}, 2013,
  \textbf{87}, 155304\relax
\mciteBstWouldAddEndPuncttrue
\mciteSetBstMidEndSepPunct{\mcitedefaultmidpunct}
{\mcitedefaultendpunct}{\mcitedefaultseppunct}\relax
\EndOfBibitem
\bibitem[Azhikodan \emph{et~al.}(2016)Azhikodan, Nautiyal, Shallcross, and
  Sharma]{bse4}
D.~Azhikodan, T.~Nautiyal, S.~Shallcross and S.~Sharma, \emph{Sci. Rep.}, 2016,
  \textbf{6}, 37075\relax
\mciteBstWouldAddEndPuncttrue
\mciteSetBstMidEndSepPunct{\mcitedefaultmidpunct}
{\mcitedefaultendpunct}{\mcitedefaultseppunct}\relax
\EndOfBibitem
\bibitem[Wirtz \emph{et~al.}(2006)Wirtz, Marini, and Rubio]{bse-m-hBN0}
L.~Wirtz, A.~Marini and A.~Rubio, \emph{Phys. Rev. Lett.}, 2006, \textbf{96},
  126104\relax
\mciteBstWouldAddEndPuncttrue
\mciteSetBstMidEndSepPunct{\mcitedefaultmidpunct}
{\mcitedefaultendpunct}{\mcitedefaultseppunct}\relax
\EndOfBibitem
\bibitem[Cudazzo \emph{et~al.}(2016)Cudazzo, Sponza, Giorgetti, Reining,
  Sottile, and Gatti]{bse-m-hBN1}
P.~Cudazzo, L.~Sponza, C.~Giorgetti, L.~Reining, F.~Sottile and M.~Gatti,
  \emph{Phys. Rev. Lett.}, 2016, \textbf{116}, 066803\relax
\mciteBstWouldAddEndPuncttrue
\mciteSetBstMidEndSepPunct{\mcitedefaultmidpunct}
{\mcitedefaultendpunct}{\mcitedefaultseppunct}\relax
\EndOfBibitem
\bibitem[Ferreira \emph{et~al.}(2019)Ferreira, Chaves, Peres, and
  Ribeiro]{bse-m-hBN2}
F.~Ferreira, A.~J. Chaves, N.~M.~R. Peres and R.~M. Ribeiro, \emph{J. Opt. Soc.
  Am. B}, 2019, \textbf{36}, 674\relax
\mciteBstWouldAddEndPuncttrue
\mciteSetBstMidEndSepPunct{\mcitedefaultmidpunct}
{\mcitedefaultendpunct}{\mcitedefaultseppunct}\relax
\EndOfBibitem
\bibitem[Ashhadi and Ketabi(2014)]{bse-bi-hBN1}
M.~Ashhadi and S.~A. Ketabi, \emph{Solid State Commun.}, 2014, \textbf{187},
  1\relax
\mciteBstWouldAddEndPuncttrue
\mciteSetBstMidEndSepPunct{\mcitedefaultmidpunct}
{\mcitedefaultendpunct}{\mcitedefaultseppunct}\relax
\EndOfBibitem
\bibitem[Paleari \emph{et~al.}(2018)Paleari, Galvani, Amara, Ducastelle,
  Molina-S{\'a}nchez, and Wirtz]{bse-bi-hBN2}
F.~Paleari, T.~Galvani, H.~Amara, F.~Ducastelle, A.~Molina-S{\'a}nchez and
  L.~Wirtz, \emph{2D Mater}, 2018, \textbf{5}, 045017\relax
\mciteBstWouldAddEndPuncttrue
\mciteSetBstMidEndSepPunct{\mcitedefaultmidpunct}
{\mcitedefaultendpunct}{\mcitedefaultseppunct}\relax
\EndOfBibitem
\bibitem[Yan \emph{et~al.}(2012)Yan, Jacobsen, and Thygesen]{bsehetero1}
J.~Yan, K.~W. Jacobsen and K.~S. Thygesen, \emph{Phys. Rev. B}, 2012,
  \textbf{86}, 045208\relax
\mciteBstWouldAddEndPuncttrue
\mciteSetBstMidEndSepPunct{\mcitedefaultmidpunct}
{\mcitedefaultendpunct}{\mcitedefaultseppunct}\relax
\EndOfBibitem
\bibitem[Komsa and Krasheninnikov(2013)]{bsehetero2}
H.-P. Komsa and A.~V. Krasheninnikov, \emph{Phys. Rev. B}, 2013, \textbf{88},
  085318\relax
\mciteBstWouldAddEndPuncttrue
\mciteSetBstMidEndSepPunct{\mcitedefaultmidpunct}
{\mcitedefaultendpunct}{\mcitedefaultseppunct}\relax
\EndOfBibitem
\bibitem[Gillen and Maultzsch(2018)]{bsehetero3}
R.~Gillen and J.~Maultzsch, \emph{Phys. Rev. B}, 2018, \textbf{97},
  165306\relax
\mciteBstWouldAddEndPuncttrue
\mciteSetBstMidEndSepPunct{\mcitedefaultmidpunct}
{\mcitedefaultendpunct}{\mcitedefaultseppunct}\relax
\EndOfBibitem
\bibitem[Runge and Gross(1984)]{tddft1}
E.~Runge and E.~K.~U. Gross, \emph{Phys. Rev. Lett.}, 1984, \textbf{52},
  997\relax
\mciteBstWouldAddEndPuncttrue
\mciteSetBstMidEndSepPunct{\mcitedefaultmidpunct}
{\mcitedefaultendpunct}{\mcitedefaultseppunct}\relax
\EndOfBibitem
\bibitem[Ullrich(2012)]{tddft2}
C.~A. Ullrich, \emph{Time-Dependent Density-Functional Theory: Concepts and
  Applications}, Oxford University Press, 2012\relax
\mciteBstWouldAddEndPuncttrue
\mciteSetBstMidEndSepPunct{\mcitedefaultmidpunct}
{\mcitedefaultendpunct}{\mcitedefaultseppunct}\relax
\EndOfBibitem
\bibitem[Maitra(2016)]{tddft3}
N.~T. Maitra, \emph{J. Chem. Phys.}, 2016, \textbf{144}, 220901\relax
\mciteBstWouldAddEndPuncttrue
\mciteSetBstMidEndSepPunct{\mcitedefaultmidpunct}
{\mcitedefaultendpunct}{\mcitedefaultseppunct}\relax
\EndOfBibitem
\bibitem[Reining \emph{et~al.}(2002)Reining, Olevano, Rubio, and Onida]{fxc1}
L.~Reining, V.~Olevano, A.~Rubio and G.~Onida, \emph{Phys. Rev. Lett.}, 2002,
  \textbf{88}, 066404\relax
\mciteBstWouldAddEndPuncttrue
\mciteSetBstMidEndSepPunct{\mcitedefaultmidpunct}
{\mcitedefaultendpunct}{\mcitedefaultseppunct}\relax
\EndOfBibitem
\bibitem[Sottile \emph{et~al.}(2003)Sottile, Olevano, and Reining]{fxc2}
F.~Sottile, V.~Olevano and L.~Reining, \emph{Phys. Rev. Lett.}, 2003,
  \textbf{91}, 056402\relax
\mciteBstWouldAddEndPuncttrue
\mciteSetBstMidEndSepPunct{\mcitedefaultmidpunct}
{\mcitedefaultendpunct}{\mcitedefaultseppunct}\relax
\EndOfBibitem
\bibitem[Marini \emph{et~al.}(2003)Marini, {R. Del Sole}, and Rubio]{fxc3}
A.~Marini, {R. Del Sole} and A.~Rubio, \emph{Phys. Rev. Lett.}, 2003,
  \textbf{91}, 256402\relax
\mciteBstWouldAddEndPuncttrue
\mciteSetBstMidEndSepPunct{\mcitedefaultmidpunct}
{\mcitedefaultendpunct}{\mcitedefaultseppunct}\relax
\EndOfBibitem
\bibitem[Botti \emph{et~al.}(2004)Botti, Sottile, Vast, Olevano, Reining,
  Weissker, Rubio, Onida, {R. Del Sole}, and Godby]{fxc4}
S.~Botti, F.~Sottile, N.~Vast, V.~Olevano, L.~Reining, H.-C. Weissker,
  A.~Rubio, G.~Onida, {R. Del Sole} and R.~W. Godby, \emph{Phys. Rev. B}, 2004,
  \textbf{69}, 155112\relax
\mciteBstWouldAddEndPuncttrue
\mciteSetBstMidEndSepPunct{\mcitedefaultmidpunct}
{\mcitedefaultendpunct}{\mcitedefaultseppunct}\relax
\EndOfBibitem
\bibitem[Botti \emph{et~al.}(2005)Botti, Fourreau, Nguyen, Renault, Sottile,
  and Reining]{fxc5}
S.~Botti, A.~Fourreau, F.~Nguyen, Y.-O. Renault, F.~Sottile and L.~Reining,
  \emph{Phys. Rev. B}, 2005, \textbf{72}, 125203\relax
\mciteBstWouldAddEndPuncttrue
\mciteSetBstMidEndSepPunct{\mcitedefaultmidpunct}
{\mcitedefaultendpunct}{\mcitedefaultseppunct}\relax
\EndOfBibitem
\bibitem[Sharma \emph{et~al.}(2011)Sharma, Dewhurst, Sanna, and Gross]{boot1}
S.~Sharma, J.~K. Dewhurst, A.~Sanna and E.~K.~U. Gross, \emph{Phys. Rev.
  Lett.}, 2011, \textbf{107}, 186401\relax
\mciteBstWouldAddEndPuncttrue
\mciteSetBstMidEndSepPunct{\mcitedefaultmidpunct}
{\mcitedefaultendpunct}{\mcitedefaultseppunct}\relax
\EndOfBibitem
\bibitem[Yang \emph{et~al.}(2012)Yang, Li, and Ullrich]{fxc6}
Z.-H. Yang, Y.~Li and C.~A. Ullrich, \emph{J. Chem. Phys.}, 2012, \textbf{137},
  014513\relax
\mciteBstWouldAddEndPuncttrue
\mciteSetBstMidEndSepPunct{\mcitedefaultmidpunct}
{\mcitedefaultendpunct}{\mcitedefaultseppunct}\relax
\EndOfBibitem
\bibitem[Yang and Ullrich(2013)]{fxc7}
Z.-H. Yang and C.~A. Ullrich, \emph{Phys. Rev. B}, 2013, \textbf{87},
  195204\relax
\mciteBstWouldAddEndPuncttrue
\mciteSetBstMidEndSepPunct{\mcitedefaultmidpunct}
{\mcitedefaultendpunct}{\mcitedefaultseppunct}\relax
\EndOfBibitem
\bibitem[Rigamonti \emph{et~al.}(2015)Rigamonti, Botti, Veniard, Draxl,
  Reining, and Sottile]{fxc8}
S.~Rigamonti, S.~Botti, V.~Veniard, C.~Draxl, L.~Reining and F.~Sottile,
  \emph{Phys. Rev. Lett.}, 2015, \textbf{114}, 146402\relax
\mciteBstWouldAddEndPuncttrue
\mciteSetBstMidEndSepPunct{\mcitedefaultmidpunct}
{\mcitedefaultendpunct}{\mcitedefaultseppunct}\relax
\EndOfBibitem
\bibitem[Yang \emph{et~al.}(2015)Yang, Sottile, and Ullrich]{fxc9}
Z.-H. Yang, F.~Sottile and C.~A. Ullrich, \emph{Phys. Rev. B}, 2015,
  \textbf{92}, 035202\relax
\mciteBstWouldAddEndPuncttrue
\mciteSetBstMidEndSepPunct{\mcitedefaultmidpunct}
{\mcitedefaultendpunct}{\mcitedefaultseppunct}\relax
\EndOfBibitem
\bibitem[Berger(2015)]{fxc92}
J.~A. Berger, \emph{Phys. Rev. Lett.}, 2015, \textbf{115}, 137402\relax
\mciteBstWouldAddEndPuncttrue
\mciteSetBstMidEndSepPunct{\mcitedefaultmidpunct}
{\mcitedefaultendpunct}{\mcitedefaultseppunct}\relax
\EndOfBibitem
\bibitem[Byun and Ullrich(2017)]{fxc10}
Y.-M. Byun and C.~A. Ullrich, \emph{Phys. Rev. B}, 2017, \textbf{95},
  205136\relax
\mciteBstWouldAddEndPuncttrue
\mciteSetBstMidEndSepPunct{\mcitedefaultmidpunct}
{\mcitedefaultendpunct}{\mcitedefaultseppunct}\relax
\EndOfBibitem
\bibitem[Byun and Ullrich(2017)]{fxc11}
Y.-M. Byun and C.~A. Ullrich, \emph{Computation}, 2017, \textbf{5}, 9\relax
\mciteBstWouldAddEndPuncttrue
\mciteSetBstMidEndSepPunct{\mcitedefaultmidpunct}
{\mcitedefaultendpunct}{\mcitedefaultseppunct}\relax
\EndOfBibitem
\bibitem[Terentjev \emph{et~al.}(2018)Terentjev, Constantin, and
  Pitarke]{fxc12}
A.~V. Terentjev, L.~A. Constantin and J.~M. Pitarke, \emph{Phys. Rev. B}, 2018,
  \textbf{98}, 085123\relax
\mciteBstWouldAddEndPuncttrue
\mciteSetBstMidEndSepPunct{\mcitedefaultmidpunct}
{\mcitedefaultendpunct}{\mcitedefaultseppunct}\relax
\EndOfBibitem
\bibitem[Singh \emph{et~al.}(2019)Singh, Elliott, Nautiyal, Dewhurst, and
  Sharma]{fxc13}
N.~Singh, P.~Elliott, T.~Nautiyal, J.~K. Dewhurst and S.~Sharma, \emph{Phys.
  Rev. B}, 2019, \textbf{99}, 035151\relax
\mciteBstWouldAddEndPuncttrue
\mciteSetBstMidEndSepPunct{\mcitedefaultmidpunct}
{\mcitedefaultendpunct}{\mcitedefaultseppunct}\relax
\EndOfBibitem
\bibitem[Sharma \emph{et~al.}(2015)Sharma, Dewhurst, Shallcross, Madjarova, and
  Gross]{boot2}
S.~Sharma, J.~K. Dewhurst, S.~Shallcross, G.~K. Madjarova and E.~K.~U. Gross,
  \emph{J. Chem. Theory Comput.}, 2015, \textbf{11}, 1710\relax
\mciteBstWouldAddEndPuncttrue
\mciteSetBstMidEndSepPunct{\mcitedefaultmidpunct}
{\mcitedefaultendpunct}{\mcitedefaultseppunct}\relax
\EndOfBibitem
\bibitem[Tran and Blaha(2009)]{tb1}
F.~Tran and P.~Blaha, \emph{Phys. Rev. Lett,}, 2009, \textbf{102}, 226401\relax
\mciteBstWouldAddEndPuncttrue
\mciteSetBstMidEndSepPunct{\mcitedefaultmidpunct}
{\mcitedefaultendpunct}{\mcitedefaultseppunct}\relax
\EndOfBibitem
\bibitem[Singh(2010)]{tb2}
D.~J. Singh, \emph{Phys. Rev. B}, 2010, \textbf{82}, 205102\relax
\mciteBstWouldAddEndPuncttrue
\mciteSetBstMidEndSepPunct{\mcitedefaultmidpunct}
{\mcitedefaultendpunct}{\mcitedefaultseppunct}\relax
\EndOfBibitem
\bibitem[Sato \emph{et~al.}(2015)Sato, Taniguchi, Shinohara, and Yabana]{tb3}
S.~A. Sato, Y.~Taniguchi, Y.~Shinohara and K.~Yabana, \emph{J. Chem. Phys.},
  2015, \textbf{143}, 224116\relax
\mciteBstWouldAddEndPuncttrue
\mciteSetBstMidEndSepPunct{\mcitedefaultmidpunct}
{\mcitedefaultendpunct}{\mcitedefaultseppunct}\relax
\EndOfBibitem
\bibitem[Li \emph{et~al.}(2017)Li, Ren, and He]{tddft2D1}
P.~Li, X.~Ren and L.~He, \emph{Phys. Rev. B}, 2017, \textbf{96}, 165417\relax
\mciteBstWouldAddEndPuncttrue
\mciteSetBstMidEndSepPunct{\mcitedefaultmidpunct}
{\mcitedefaultendpunct}{\mcitedefaultseppunct}\relax
\EndOfBibitem
\bibitem[Turkowski \emph{et~al.}(2017)Turkowski, Din, and Rahman]{tddft2D2}
V.~Turkowski, N.~U. Din and T.~S. Rahman, \emph{Computation}, 2017, \textbf{5},
  39\relax
\mciteBstWouldAddEndPuncttrue
\mciteSetBstMidEndSepPunct{\mcitedefaultmidpunct}
{\mcitedefaultendpunct}{\mcitedefaultseppunct}\relax
\EndOfBibitem
\bibitem[Yamada \emph{et~al.}(2018)Yamada, Noda, Nobusada, and
  Yabana]{tddft2D3}
S.~Yamada, M.~Noda, K.~Nobusada and K.~Yabana, \emph{Phys. Rev. B}, 2018,
  \textbf{98}, 245147\relax
\mciteBstWouldAddEndPuncttrue
\mciteSetBstMidEndSepPunct{\mcitedefaultmidpunct}
{\mcitedefaultendpunct}{\mcitedefaultseppunct}\relax
\EndOfBibitem
\bibitem[Miyamoto and Rubio(2018)]{tddft2D4}
Y.~Miyamoto and A.~Rubio, \emph{J. Phys. Soc. Jpn.}, 2018, \textbf{87},
  041016\relax
\mciteBstWouldAddEndPuncttrue
\mciteSetBstMidEndSepPunct{\mcitedefaultmidpunct}
{\mcitedefaultendpunct}{\mcitedefaultseppunct}\relax
\EndOfBibitem
\bibitem[{U. De Giovannini} \emph{et~al.}(2017){U. De Giovannini}, H{\"u}bener,
  and Rubio]{tddft2D5}
{U. De Giovannini}, H.~H{\"u}bener and A.~Rubio, \emph{J. Chem. Theory
  Comput.}, 2017, \textbf{13}, 265\relax
\mciteBstWouldAddEndPuncttrue
\mciteSetBstMidEndSepPunct{\mcitedefaultmidpunct}
{\mcitedefaultendpunct}{\mcitedefaultseppunct}\relax
\EndOfBibitem
\bibitem[Sato \emph{et~al.}(2018)Sato, H{\"u}bener, {U. De Giovannini}, and
  Rubio]{tddft2D6}
S.~A. Sato, H.~H{\"u}bener, {U. De Giovannini} and A.~Rubio, \emph{Appl. Sci.},
  2018, \textbf{8}, 1777\relax
\mciteBstWouldAddEndPuncttrue
\mciteSetBstMidEndSepPunct{\mcitedefaultmidpunct}
{\mcitedefaultendpunct}{\mcitedefaultseppunct}\relax
\EndOfBibitem
\bibitem[Wang \emph{et~al.}(2015)Wang, Li, and Wang]{tddft2D7}
Z.~Wang, S.-S. Li and L.-W. Wang, \emph{Phys. Rev. Lett.}, 2015, \textbf{114},
  063004\relax
\mciteBstWouldAddEndPuncttrue
\mciteSetBstMidEndSepPunct{\mcitedefaultmidpunct}
{\mcitedefaultendpunct}{\mcitedefaultseppunct}\relax
\EndOfBibitem
\bibitem[Ueda \emph{et~al.}(2018)Ueda, Suzuki, and Watanabe]{tddft2D8}
Y.~Ueda, Y.~Suzuki and K.~Watanabe, \emph{Phys. Rev. B}, 2018, \textbf{97},
  075406\relax
\mciteBstWouldAddEndPuncttrue
\mciteSetBstMidEndSepPunct{\mcitedefaultmidpunct}
{\mcitedefaultendpunct}{\mcitedefaultseppunct}\relax
\EndOfBibitem
\bibitem[Aggoune \emph{et~al.}(2018)Aggoune, Cocchi, Nabok, Rezouali, Belkhir,
  and Draxl]{bse-bulk-hBN}
W.~Aggoune, C.~Cocchi, D.~Nabok, K.~Rezouali, M.~A. Belkhir and C.~Draxl,
  \emph{Phys. Rev. B}, 2018, \textbf{97}, 241114(R)\relax
\mciteBstWouldAddEndPuncttrue
\mciteSetBstMidEndSepPunct{\mcitedefaultmidpunct}
{\mcitedefaultendpunct}{\mcitedefaultseppunct}\relax
\EndOfBibitem
\bibitem[Li \emph{et~al.}(2014)Li, Chernikov, Zhang, Rigosi, Hill, {A. M. van
  der Zande}, Chenet, Shih, Hone, and Heinz]{tmdcexp}
Y.~Li, A.~Chernikov, X.~Zhang, A.~Rigosi, H.~M. Hill, {A. M. van der Zande},
  D.~A. Chenet, E.-M. Shih, J.~Hone and T.~F. Heinz, \emph{Phys. Rev. B}, 2014,
  \textbf{90}, 205422\relax
\mciteBstWouldAddEndPuncttrue
\mciteSetBstMidEndSepPunct{\mcitedefaultmidpunct}
{\mcitedefaultendpunct}{\mcitedefaultseppunct}\relax
\EndOfBibitem
\bibitem[Stubner \emph{et~al.}(2004)Stubner, Tokatly, and Pankratov]{fxcqp}
R.~Stubner, I.~V. Tokatly and O.~Pankratov, \emph{Phys. Rev. B}, 2004,
  \textbf{70}, 245119\relax
\mciteBstWouldAddEndPuncttrue
\mciteSetBstMidEndSepPunct{\mcitedefaultmidpunct}
{\mcitedefaultendpunct}{\mcitedefaultseppunct}\relax
\EndOfBibitem
\bibitem[{Ph. Ghosez} \emph{et~al.}(1997){Ph. Ghosez}, Gonze, and Godby]{div1}
{Ph. Ghosez}, X.~Gonze and R.~W. Godby, \emph{Phys. Rev. B}, 1997, \textbf{56},
  12811\relax
\mciteBstWouldAddEndPuncttrue
\mciteSetBstMidEndSepPunct{\mcitedefaultmidpunct}
{\mcitedefaultendpunct}{\mcitedefaultseppunct}\relax
\EndOfBibitem
\bibitem[Kim and G{\"o}rling(2002)]{div2}
Y.-H. Kim and A.~G{\"o}rling, \emph{Phys. Rev. B}, 2002, \textbf{66},
  035114\relax
\mciteBstWouldAddEndPuncttrue
\mciteSetBstMidEndSepPunct{\mcitedefaultmidpunct}
{\mcitedefaultendpunct}{\mcitedefaultseppunct}\relax
\EndOfBibitem
\bibitem[Maitra(2017)]{EXX}
N.~T. Maitra, \emph{J. Phys.: Condens. Matter}, 2017, \textbf{29}, 423001\relax
\mciteBstWouldAddEndPuncttrue
\mciteSetBstMidEndSepPunct{\mcitedefaultmidpunct}
{\mcitedefaultendpunct}{\mcitedefaultseppunct}\relax
\EndOfBibitem
\bibitem[Levine and Allan(1991)]{scissor}
Z.~H. Levine and D.~C. Allan, \emph{Phys. Rev. B}, 1991, \textbf{43},
  4187\relax
\mciteBstWouldAddEndPuncttrue
\mciteSetBstMidEndSepPunct{\mcitedefaultmidpunct}
{\mcitedefaultendpunct}{\mcitedefaultseppunct}\relax
\EndOfBibitem
\bibitem[elk(accessed 2017)]{elk}
\emph{The Elk FP-LAPW Code}, accessed 2017\relax
\mciteBstWouldAddEndPuncttrue
\mciteSetBstMidEndSepPunct{\mcitedefaultmidpunct}
{\mcitedefaultendpunct}{\mcitedefaultseppunct}\relax
\EndOfBibitem
\bibitem[Gulans \emph{et~al.}(2014)Gulans, Kontur, Meisenbichler, Nabok,
  Pavone, Rigamonti, Sagmeister, Werner, and Draxl]{exciting1}
A.~Gulans, S.~Kontur, C.~Meisenbichler, D.~Nabok, P.~Pavone, S.~Rigamonti,
  S.~Sagmeister, U.~Werner and C.~Draxl, \emph{J. Phys.: Condens. Matter.},
  2014, \textbf{26}, 363202\relax
\mciteBstWouldAddEndPuncttrue
\mciteSetBstMidEndSepPunct{\mcitedefaultmidpunct}
{\mcitedefaultendpunct}{\mcitedefaultseppunct}\relax
\EndOfBibitem
\bibitem[Sagmeister and Draxl(2009)]{exciting2}
S.~Sagmeister and C.~Draxl, \emph{Phys. Chem. Chem. Phys.}, 2009, \textbf{11},
  4451\relax
\mciteBstWouldAddEndPuncttrue
\mciteSetBstMidEndSepPunct{\mcitedefaultmidpunct}
{\mcitedefaultendpunct}{\mcitedefaultseppunct}\relax
\EndOfBibitem
\bibitem[Nabok \emph{et~al.}(2016)Nabok, Gulans, and Draxl]{exciting3}
D.~Nabok, A.~Gulans and C.~Draxl, \emph{Phys. Rev. B}, 2016, \textbf{94},
  035118\relax
\mciteBstWouldAddEndPuncttrue
\mciteSetBstMidEndSepPunct{\mcitedefaultmidpunct}
{\mcitedefaultendpunct}{\mcitedefaultseppunct}\relax
\EndOfBibitem
\bibitem[Perdew \emph{et~al.}(1996)Perdew, Burke, and Ernzerhof]{pbe}
J.~P. Perdew, K.~Burke and M.~Ernzerhof, \emph{Phys. Rev. Lett.}, 1996,
  \textbf{77}, 3865\relax
\mciteBstWouldAddEndPuncttrue
\mciteSetBstMidEndSepPunct{\mcitedefaultmidpunct}
{\mcitedefaultendpunct}{\mcitedefaultseppunct}\relax
\EndOfBibitem
\bibitem[Arnaud \emph{et~al.}(2006)Arnaud, Lebegue, Rabiller, and
  Alouani]{bse-bulk-hBN2}
B.~Arnaud, S.~Lebegue, P.~Rabiller and M.~Alouani, \emph{Phys. Rev. Lett.},
  2006, \textbf{96}, 026402\relax
\mciteBstWouldAddEndPuncttrue
\mciteSetBstMidEndSepPunct{\mcitedefaultmidpunct}
{\mcitedefaultendpunct}{\mcitedefaultseppunct}\relax
\EndOfBibitem
\bibitem[Wirtz \emph{et~al.}(2008)Wirtz, Marini, Gr{\"u}ning, Attaccalite,
  Kresse, and Rubio]{bse-bulk-hBN3}
L.~Wirtz, A.~Marini, M.~Gr{\"u}ning, C.~Attaccalite, G.~Kresse and A.~Rubio,
  \emph{Phys. Rev. Lett.}, 2008, \textbf{100}, 189701\relax
\mciteBstWouldAddEndPuncttrue
\mciteSetBstMidEndSepPunct{\mcitedefaultmidpunct}
{\mcitedefaultendpunct}{\mcitedefaultseppunct}\relax
\EndOfBibitem
\bibitem[Liu \emph{et~al.}(2003)Liu, Feng, and Shen]{bse-bulk-hBN4}
L.~Liu, Y.~P. Feng and Z.~X. Shen, \emph{Phys. Rev. B}, 2003, \textbf{68},
  104102\relax
\mciteBstWouldAddEndPuncttrue
\mciteSetBstMidEndSepPunct{\mcitedefaultmidpunct}
{\mcitedefaultendpunct}{\mcitedefaultseppunct}\relax
\EndOfBibitem
\bibitem[Gao(2012)]{bse-bulk-hBN5}
S.-P. Gao, \emph{Solid State Commun.}, 2012, \textbf{152}, 1817\relax
\mciteBstWouldAddEndPuncttrue
\mciteSetBstMidEndSepPunct{\mcitedefaultmidpunct}
{\mcitedefaultendpunct}{\mcitedefaultseppunct}\relax
\EndOfBibitem
\bibitem[Ismail-Beigi(2006)]{truncation}
S.~Ismail-Beigi, \emph{Phys. Rev. B}, 2006, \textbf{73}, 233103\relax
\mciteBstWouldAddEndPuncttrue
\mciteSetBstMidEndSepPunct{\mcitedefaultmidpunct}
{\mcitedefaultendpunct}{\mcitedefaultseppunct}\relax
\EndOfBibitem
\bibitem[Perdew and Wang(1992)]{lda}
J.~P. Perdew and Y.~Wang, \emph{Phys. Rev. B}, 1992, \textbf{45}, 13244\relax
\mciteBstWouldAddEndPuncttrue
\mciteSetBstMidEndSepPunct{\mcitedefaultmidpunct}
{\mcitedefaultendpunct}{\mcitedefaultseppunct}\relax
\EndOfBibitem
\bibitem[Chernikov \emph{et~al.}(2014)Chernikov, Berkelbach, Hill, Rigosi, Li,
  Aslan, Reichman, Hybertsen, and Heinz]{rydberg1}
A.~Chernikov, T.~C. Berkelbach, H.~M. Hill, A.~Rigosi, Y.~Li, O.~B. Aslan,
  D.~R. Reichman, M.~S. Hybertsen and T.~F. Heinz, \emph{Phys. Rev. Lett.},
  2014, \textbf{113}, 076802\relax
\mciteBstWouldAddEndPuncttrue
\mciteSetBstMidEndSepPunct{\mcitedefaultmidpunct}
{\mcitedefaultendpunct}{\mcitedefaultseppunct}\relax
\EndOfBibitem
\bibitem[Bergh{\"a}user and Malic(2014)]{rydberg2}
G.~Bergh{\"a}user and E.~Malic, \emph{Phys. Rev. B}, 2014, \textbf{89},
  125309\relax
\mciteBstWouldAddEndPuncttrue
\mciteSetBstMidEndSepPunct{\mcitedefaultmidpunct}
{\mcitedefaultendpunct}{\mcitedefaultseppunct}\relax
\EndOfBibitem
\end{mcitethebibliography}
\bibliographystyle{rsc} 

\end{document}